\documentclass[letterpaper,times]{IONconf}
\usepackage{url}

\usepackage[colorlinks,urlcolor=blue,linkcolor=blue,citecolor=blue,breaklinks=true]{hyperref}

\usepackage{siunitx}
\usepackage{lipsum}
\usepackage{amsmath}
\usepackage{amssymb}
\usepackage{amsfonts}

\usepackage{bm}
\usepackage{tabularray}
\usepackage{graphicx}
\usepackage{float}
\usepackage{subfig}
\usepackage{xcolor}
\usepackage{environ}
\NewEnviron{review}{\textcolor{black}{\BODY}}
\NewEnviron{dummy}{\textcolor[rgb]{0.8,0.8,0.8}{\BODY}}
\usepackage{threeparttable}
\usepackage{csquotes}
\newcommand*\diff{\mathop{}\!\mathrm{d}}
\newcommand{\keywords}[1]{\par\addvspace\baselineskip
\noindent\textbf{Keywords:} #1\par}

\title{Jackknife ARAIM: Efficient GNSS Integrity Monitoring for Simultaneous Faults under Non-Gaussian Errors}

\author{
    Penggao Yan \\
    \textit{Department of Aeronautical and Aviation Engineering, The Hong Kong Polytechnic University}\\
    Ronghe Jin \\
    \textit{School of Remote Sensing and Information Engineering, Wuhan University}\\
    Junyi Zhang \\
    \textit{Department of Applied Mathematics, The Hong Kong Polytechnic University}\\
    Cheng-Wei Wang \\
    \textit{Department of Aeronautics and Astronautics, National Cheng Kung University}\\
    Li-Ta Hsu* \\
    \textit{Department of Aeronautical and Aviation Engineering, The Hong Kong Polytechnic University}\\
Email: lt.hsu@polyu.edu.hk
    }

\begin{document}

\maketitle

\pagestyle{myheadings}
\setcounter{page}{1}








\section*{Abstract}
Global navigation satellite systems (GNSS) require strict integrity monitoring for aviation safety, where legacy and advanced receiver autonomous integrity monitoring (RAIM/ARAIM) rely on Gaussian error models that can be overly conservative for real-world non-Gaussian errors.
This paper proposes an extended jackknife detector capable of detecting multiple simultaneous faults with non-Gaussian nominal errors.
\begin{review}The proposed detector adopts a hybrid monitoring strategy: scalar jackknife residual tests for single-fault modes and triple-axis (East/North/Up) tests for multi-fault modes\end{review}.
Furthermore, an integrity monitoring algorithm, jackknife ARAIM, is developed by systematically exploiting the properties of the jackknife detector in the range domain.
\begin{review}We prove that the proposed method has equivalent monitoring performance with the \textcolor{black}{solution separation (SS) ARAIM}, but is significantly computationally efficient for single-fault cases with non-Gaussian nominal errors, while maintaining similar efficiency to \textcolor{black}{SS ARAIM} for multiple-fault cases\end{review}.
The proposed method is examined in worldwide simulations, with the nominal measurement error simulated based on authentic experimental data, which reveals different findings in existing research.
In a single Global Positioning System (GPS) constellation setting, the proposed method can reduce the 99.5 percentile vertical protection level (VPL) below \SI{45}{\meter}, \begin{review}outperforming \SI{50}{\meter} VPL produced by the ARAIM algorithm using Gaussian nominal error models\end{review}.
In GPS-Galileo dual-constellation setting, while these Gaussian-based ARAIM algorithms suffer VPL inflation over \SI{60}{\meter} due to Galileo's heavy-tailed errors, the proposed method maintains VPL below \SI{40}{\meter}, achieving over \SI{92}{\percent} normal operations for \SI{35}{\meter} Vertical Alert Limit.
Moreover, we tentatively implement the \textcolor{black}{SS ARAIM} using non-Gaussian overbounds and compare it with the proposed Jackknife ARAIM method regarding computation efficiency.
\begin{review}The proposed method achieves up to \SI{59.4}{\percent} reduction in median processing time compared to \textcolor{black}{SS ARAIM} in single-constellation scenarios. In dual-constellation scenarios, where multiple-fault cases are more prevalent, both algorithms exhibit similar computational efficiency\end{review}.
These improvements enable potential support for localizer performance with vertical guidance (LPV) with a decision height of 200 ft, enhancing integrity and computation efficiency for multi-constellation GNSS applications.

\keywords{Receiver autonomous integrity monitoring, Fault detection, Heavy-tailed errors, Simultaneous faults, Global navigation satellite system}


\section{Introduction}
Global navigation satellite systems (GNSS) have become indispensable for positioning and navigation in safety-critical applications such as aviation, where strict integrity requirements must be met \citep{brown_baseline_1992,perea2017ura,blanch2022baseline,yan2025fault,zhai2020integrated}.
Integrity monitoring ensures that users are alerted when GNSS positioning errors exceed tolerable limits, preventing hazardously misleading information \citep{WGC2015}.
In the era of Global Positioning System (GPS) navigation, this has been achieved through receiver autonomous integrity monitoring (RAIM), which uses redundant pseudorange measurements to detect faults by consistency checks \citep{parkinson_autonomous_1988,brown_baseline_1992}.
RAIM has been widely adopted due to its ability to provide integrity alerts autonomously on the receiver side.
However, legacy RAIM has inherent limitations: it typically assumes at most one faulty satellite at a time and relies on a single constellation, providing only horizontal-plane protection and no guaranteed vertical guidance \citep{angus_raim_2006,pervan_multiple_1998,blanch2015baseline}.
These constraints mean that legacy RAIM may become unavailable or ineffective in complex scenarios (e.g., multiple simultaneous satellite faults or degraded geometry \citep{hutsell2002one,blanch2015baseline}), prompting the need for more advanced integrity solutions.

Advanced RAIM (ARAIM) was proposed to overcome the above limitations by leveraging multi-constellation and dual-frequency GNSS measurements \citep{blanch2015baseline,joerger_solution_2014}.
By incorporating signals from multiple GNSS and using fault-tolerant algorithms, ARAIM can handle more complex fault hypotheses (including multiple simultaneous satellite faults) at the cost of higher computational load.
This allows ARAIM to improve service availability and support vertical navigation integrity in a multi-GNSS environment \citep{perea2017ura,blanch2013critical,icao2006Annex10}.
ARAIM is being developed with the goal of enabling worldwide precision approach operations, such as localizer performance with vertical guidance (LPV) with a decision height of 200 ft, without the need for local augmentation systems. 
\begin{review}
Beyond its original pseudorange-based positioning framework, recent research has extended ARAIM to incorporate carrier-phase measurements through precise point positioning techniques \citep{wang2025integrity,zhang2024gnss}, enabling higher precision positioning while maintaining integrity guarantees. Furthermore, ARAIM has been adapted for applications beyond aviation, such as unmanned aerial vehicles \citep{moreno2024experimentation} and rail \citep{neri2025ran}.
\end{review}

Despite these advancements, a common assumption has been made in developing RAIM and ARAIM algorithms that the nominal measurement error is Gaussian distributed, which simplifies the derivation and reduces the computational effort.
However, nominal range errors in the real world usually have non-Gaussian and heavy-tailed properties \citep{rife_core_2004,braff2005method,pervan2000sigma,chang2021analysis}.
For example, as important components of range errors, orbit and clock errors show significantly heavy-tailed properties \citep{perea2017ura,wang2021characterizing,wu2020analysis,huo2023integrity}, making their Gaussian overbound over-conservative.
Such over-conservatism will be passed to the position domain and enlarge the protection level (PL) of the ARAIM algorithm, eventually hindering the system's availability in real-world applications under stringent navigation requirements, such as the LPV-200 precision approach \citep{icao2006Annex10}.
The Gaussian assumption shortfall motivates the development of advanced fault detection methods that can reliably identify measurement faults or outliers with non-Gaussian errors.\par

Several approaches have been explored to address non-Gaussian errors in GNSS integrity monitoring.
\begin{review}One approach is to improve the statistical error models used for integrity monitoring \citep{rife_core_2004,larson_gaussianpareto_2019, blanch_position_2008,yan2024principal,li2025paired}. For example, \citet{gao2025enhanced} propose to use the Student's t-distribution to bound non-Gaussian errors in GNSS positioning. \citet{alghananim2025maximum} propose the maximum non-bounded difference method to effectively overbound non-Gaussian distributions. Instead of a single model, researchers have applied mixture models \citep{blanch_position_2008,yan2024principal} and extreme value theory \citep{larson_gaussianpareto_2019} to model the distribution of GNSS errors with greater fidelity to empirical data\end{review}.
In the author's previous work \citep{yan2024principal}, a non-Gaussian overbounding method (Principal Gaussian overbound) is developed by leveraging a Gaussian mixture to capture the tails of the error distribution while still providing analytically tractable bounds on integrity risk. 
However, such error bounding techniques alone do not directly pinpoint which measurements are faulty; they mainly ensure that the error magnitude can be tightly bounded \citep{yan2024principal,larson_gaussianpareto_2019,rife2012overbounding}.
Another line of work focuses on robust estimation and detection \citep{pfeifer2019expectation,yang2016gnss,garcia2020design,li2023improved}.
Robust statistical estimators (e.g., M-estimators or other regression techniques) have been introduced to lessen the influence of outliers on the position solution \citep{garcia2020design, yang2016gnss}.
These methods effectively down-weight or exclude measurements that appear inconsistent, thus improving positioning performance under heavy-tailed errors.
Yet, a challenge with many robust techniques is ensuring rigorous integrity guarantees: it can be difficult to quantitatively bound the probabilities of missed detection and false alarm without a clear underlying statistical test.

In our previous work \citep{yan2025efficient}, we introduced a fault detection method based on the jackknife statistical resampling technique to handle non-Gaussian measurement errors.
The jackknife-based GNSS fault detector operates by systematically leaving out one measurement at a time and examining the inconsistency between the observed measurement and the predicted measurement based on subset solutions.
Unlike conservative bounding methods that simply inflate error margins \citep{pervan2000sigma} or black-box robust algorithms that lack transparency \citep{crespillo2018tightly,pfeifer2019expectation}, the jackknife detector is grounded in a linearized model of the GNSS solution and yields a provably sensitive and reliable fault test \citep{yan2025efficient}.
We prove the theoretical equivalence between the jackknife detector and the solution separation detector, where the solution separation detector is the projection of the jackknife test detector along the direction defined by the derivative of the solution with respect to the suspected measurement.
Moreover, we also demonstrate that the jackknife detector yields a \begin{review}threefold~\end{review} enhancement in computational efficiency compared to the solution separation detector due to its scalar nature.
However, a significant limitation remains: the jackknife detector was developed under the assumption of a single faulty measurement per epoch, aligning with the legacy RAIM paradigm.
This restriction severely limits its applicability in real-world scenarios where multiple simultaneous faults may occur.

In this paper, we address this limitation by extending the jackknife detector to handle simultaneous faults while preserving its robustness to non-Gaussian nominal errors in linearized pseudorange-based positioning systems.
Leveraging this enhanced detection capability, we develop an integrity monitoring algorithm called jackknife ARAIM that integrates the extended jackknife detector into a multiple-hypothesis framework.
This integration enables efficient handling of both Gaussian and non-Gaussian nominal error bounds while effectively managing the computational complexity inherent in multi-fault scenarios.
While the \textcolor{black}{SS ARAIM} algorithm can accommodate non-Gaussian errors \citep{perea2019multi}, our approach inherits the efficient detection capability from the jackknife detector, enabling more computationally efficient simultaneous-fault monitoring compared to \textcolor{black}{SS ARAIM}.
To validate the practical applicability of our method, we conduct worldwide simulations using authentic experimental data to model nominal measurement errors, focusing particularly on the challenging LPV-200 precision approach requirements in a GPS-Galileo dual constellation environment. The use of authentic experimental data enhances the reliability of the experimental results. \begin{review}The contribution of this work is to extend the jackknife detector for simultaneous fault detection in linearized pseudorange-based positioning systems under non-Gaussian nominal errors and develop a computationally efficient multi-hypothesis integrity monitoring method with non-Gaussian errors.\end{review} \par

The remaining part of this paper is organized as follows.
Section \ref{sec:multiple_JK} gives a brief review of the single-fault jackknife detector and then extends this method for simultaneous fault detection.
In Section \ref{sec:JackRAIM}, a novel integrity monitoring algorithm, jackknife ARAIM, is developed by systematically exploiting the properties of the jackknife detector.
Section \ref{sec:simulation} evaluates the integriy monitoring performance of the jackknife ARAIM in a worldwide simulation with both single and dual constellations, while Section \ref{sec:computation} evaluates the computation efficiency of the proposed method.
Finally, Section \ref{sec_conclusions} presents a summary.


\section{Jackknife Detector for Simultaneous Faults}\label{sec:multiple_JK}
\subsection{Jackknife Detector for Single Fault}
In our previous work, we developed a fault detection method based on the jackknife technique, referred to as the jackknife detector, to identify faulty GNSS measurements with non-Gaussian measurement errors \citep{yan2025efficient}.
The jackknife detector computes the inconsistency between the observed and predicted measurements, which are derived from subset solutions, and performs multiple tests to identify faults.
The jackknife detector shares the common logic of solution separation \citep{pervan_multiple_1998,blanch_advanced_2012} to compute the full set and subset solutions.
Our previous work \citep{yan2025efficient} proves the theoretical equivalence between the jackknife detector and solution separation detector, where the solution separation detector is the projection of the jackknife test detector along the direction defined by the derivative of the solution with respect to the suspected measurement.
More importantly, the jackknife detector is proven to yield a \begin{review}threefold~\end{review} enhancement in computational efficiency compared to the SS detector \citep{yan2025efficient}.
A brief introduction to the jackknife detector is given as follows:\par

Using the conventions in \citet{yan2025efficient}, a generalized linear system for GNSS positioning can be written as 
\begin{equation}
    \mathbf{y} = \mathbf{G}\mathbf{x}+\bm{\varepsilon} \,,
    \label{eq:general_linear}
\end{equation}
where
\begin{equation}
\begin{aligned}
    \mathbf{y} =\begin{bmatrix}
        f\big(\rho_1,\mathbf{x}_0\big)\\
        \vdots \\
        f\big(\rho_n,\mathbf{x}_0\big)
    \end{bmatrix}\,,
    &\mathbf{G} = \begin{bmatrix}
     \mathbf{g}\big(\{p^{1,j}\},\mathbf{x}_0\big)\\
     \vdots  \\
    \mathbf{g}\big(\{p^{n,j}\},\mathbf{x}_0\big)\\
    \end{bmatrix} \,,
    \bm{\varepsilon}=\begin{bmatrix}
        \varepsilon_1 \\
        \vdots \\
        \varepsilon_n
    \end{bmatrix} \,, \\
&\mathbf{x}{=} \mathbf{x}_t - \mathbf{x}_0\,.
\label{eq:general_linear2}
\end{aligned}
\end{equation}
In the above notations, $\mathbf{x}$ is the system state (an $m\times 1$ vector); $\mathbf{x}_t$ is the receiver positioning state and $\mathbf{x}_0$ is its linearized point; $\varepsilon_i$ is the $i$th measurement error; $f\big(\rho_i,\mathbf{x}_0\big)$ is a function of the $i$th measurement $\rho_i$ (note that $\rho_i$ refers to a generalized measurement, not limited to the pseudorange measurement) and $\mathbf{x}_0$; $\mathbf{g}\big(\{p^{i,j}\},\mathbf{x}_0\big)$ is a vector function of the collection of satellite positions $\{p^{i,j}\}$ related to the $i$th measurement and $\mathbf{x}_0$; and $\mathbf{G}$ is an $n\times m$ matrix.
\par
The full set solution $\hat{\mathbf{x}}_t$ can be solved by the weighted least squares (WLS) method using all $n$ measurements
\begin{equation}
\begin{aligned}
    \hat{\mathbf{x}} &=  \mathbf{S} \mathbf{y} \\
    \hat{\mathbf{x}}_t &=  \mathbf{x}_0 + \hat{\mathbf{x}} \,,
\end{aligned}
\label{eq:full solution}
\end{equation}
where $\mathbf{S}$ is the solution matrix
\begin{equation}
\mathbf{S} = (\mathbf{G}^T\mathbf{W}\mathbf{G})^{-1}\mathbf{G}^T\mathbf{W}\,.
\end{equation}
To obtain the $k$th subset solution, the measurements with indices $i \not \in idx^{ex}_k$ are excluded.
In the jackknife detector, only a single faulty measurement is considered, i.e., $|idx^{ex}_k|=1$.
The solution matrix of the $k$th subset is given by
\begin{equation}
    \mathbf{S}^{(k)} = (\mathbf{G}^T\mathbf{W}^{(k)}\mathbf{G})^{-1}\mathbf{G}^T\mathbf{W}^{(k)} \,,
    \label{eq:S^i}
\end{equation}
where $\mathbf{W}^{(k)}$ is a diagonal matrix and is defined as 
\begin{equation}
    W^{(k)}_{i,i} = \begin{cases}
        0 & \text{if}~ i=k \\
        W_{i,i}  & \text{otherwise}
\end{cases} \,.
\end{equation}
Then, the $k$th subset solution is given by
\begin{subequations}
  \begin{align} 
  \hat{\mathbf{x}}^{(k)} {=}&  \mathbf{S}^{(k)} \mathbf{y} ~ \forall k=1\cdots n \\
  \hat{\mathbf{x}}^{(k)}_t {=}& \mathbf{x}_{0}+\hat{\mathbf{x}}^{(k)} ~ \forall k=1\cdots n \,,
  \end{align}
  \label{eq:subsolution2}
\end{subequations}
where $ \hat{\mathbf{x}}^{(k)}_t$ is the estimation of the positioning state $\mathbf{x}_t^{(k)}$ associated with the $k$th subset.

The predicted $k$th measurement with the subsolution $\hat{\mathbf{x}}^{(k)}$ is given by 
\begin{equation}
\hat{y}_k = \mathbf{g}_k \hat{\mathbf{x}}^{(k)} \,,
\label{eq:prediction}
\end{equation}
where $ \mathbf{g}_k$ is the $k$th row of $\mathbf{G}$.
The jackknife residual is given by the difference between $y_k$ and $\hat{y}_k$:
\begin{equation}
    t_k=y_k - \hat{y}_k \,,
    \label{eq:JK_residual}
\end{equation}
where $y_k$ is the $k$th element of $\mathbf{y}$.
It is proven that the jackknife residual is the linear combination of measurement errors as follows \citep{yan2025efficient}:
\begin{equation}
    t_k = \sum\limits_{j=1}^{n} \tilde{p}^{(k)}_{k,j}\varepsilon_j \,,
    \label{eq:ti_3}
\end{equation}
where $\varepsilon_j$ is the $j$th element of $\bm{\varepsilon}$, and $\tilde{p}^{(k)}_{k,j}$ is the  $(k,j)$th element of $\left(\mathbf{I}-\tilde{\mathbf{P}}^{(k)}\right)$ with $\tilde{\mathbf{P}}^{(k)}$ defined as follows:
\begin{equation}
\tilde{\mathbf{P}}^{(k)} = \mathbf{G} \mathbf{S}^{(k)} \,.
    \label{eq:P_tilde}
\end{equation}
Remarkably, $\varepsilon_j$  can have an arbitrary distribution as long as it has a probability density function (PDF) $f_{\varepsilon_j}(\cdot)$.
Since $t_k$ is the weighted sum of independent random variables with zero-mean distributions, its PDF can be easily obtained by \citep{lee_jiyun_sigma_2009}
\begin{equation}
f_{t_k}(x) = {\prod\limits_{j = 1}^{n}\left| \tilde{p}^{(k)}_{k,j} \right|^{- 1}} f_{\varepsilon_1}\left( \frac{x}{\left| \tilde{p}^{(k)}_{k,1} \right|} \right)*f_{\varepsilon_2}\left( \frac{x}{\left| \tilde{p}^{(k)}_{k,2} \right|} \right)*\ldots  * f_{\varepsilon_n}\left( \frac{x}{\left| \tilde{p}^{(k)}_{k,n} \right|} \right) \,.
    \label{eq:ti_conv}
\end{equation}

The following hypotheses with the Bonferroni correction \citep{bonferroni1936teoria} are formalized:
\begin{review}
\begin{equation}
\begin{aligned}
    H_0{:}& ~\text{No failure in the} ~ n ~ \text{measurements} \\
H_1{:}& ~\text{At least one failure in the} ~ n ~ \text{measurements} \,.
\nonumber
\end{aligned}    
\end{equation}
The hypothesis testing for fault detection can be formalized by the following.\par
\noindent \textbf{[Jackknife detector test]} $H_0$ is rejected if $Q^{-1}_{t_k}(\frac{\tau}{2n})$ at significant level of $\alpha^*$, where $Q^{-1}_{t_k}(\cdot)$ is the quantile function of the distribution of $t_k$ and $\tau$ is the upper limit of $\alpha^*$.
The probability of Type I error of the corrected test is $\alpha^*$.
The probability of Type I error of the individual test is $\alpha = \frac{\tau}{n}$.\end{review}

It is proven that solution separation detector is the projection of the jackknife test detector along the direction defined by the derivative of the solution with respect to the suspected measurement (see \ref{sec:app_SS_JK}), which also reals that the Jackknife detector can provide an more efficient approach for fault detection in non-Gaussian environments than the solution separation detector \citep{yan2025efficient}.
However, the jackknife method assumes that only one faulty measurement occurs per time epoch, which was suitable for early GNSS systems with limited satellites \citep{parkinson_autonomous_1988,pervan_multiple_1998}.
As the number of satellites and constellations grows, the probability of simultaneous faults becomes non-negligible.
For example, multiple GPS satellites experienced high L1 single-frequency range errors of up to \SI{16}{\meter} due to an erroneous ionospheric correction term between May 28 and June 2, 2002 \citep{hutsell2002one}.
This highlights the need for fault detection techniques capable of handling multiple faults \citep{blanch2015baseline}.

Indeed, researchers have proposed optimal fault detection algorithms under certain assumptions \citep{carlone2014selecting}.
These algorithms evaluate the consistency of all sets of measurements and select the best set with the highest level of consistency.
One such approach is multiple-hypothesis solution separation for multiple faults in integrity monitoring \citep{blanch2015baseline}.
In the following sections, we build upon this idea to extend the jackknife detector to handle multiple fault detection with non-Gaussian nominal errors.

\subsection{Threat Model}
The threat model defined in \citet{blanch2015baseline} is utilized to reconstruct the jackknife residual in Eq.
\eqref{eq:JK_residual} to handle the multiple-fault condition.
The threat model defines a collection of error modes that partition the whole measurement space \citep{yang2013optimal,joerger_solution_2014}.
Assuming there are $n$ measurements each uniquely numbered, the threat model is constructed by defining a set of fault modes with different prior probabilities:
\begin{itemize}
\item Fault mode 0: All measurements are nominal measurements (i.e., fault-free).
The prior probability of fault mode 0 is $P_{H_0}$.
\item Fault mode $k$: Measurements with indices $k \in idx^{ex}_k$ are faulty measurements (including single or multiple faults), while measurements with indices $k \not \in idx^{ex}_k$ are nominal measurements.
The prior probability of fault mode $k$ is $P_{H_k}$.
\end{itemize}
In the above definition, the size of $idx^{ex}_k$ is the number of simultaneous faults associated with the fault mode $k$, which takes a value from $1$ to $n$.
The total number of fault modes is assumed to be $N_\text{fault modes}+1$.
Theoretically,
\begin{equation}
    N_\text{fault modes}=\sum_{k=1}^{n-k_\text{max}}\binom{n}{k}\,,
    \label{eq:Nfaultmodes}
\end{equation}
where $k_\text{max}$ is the maximum number of simultaneous faults that need to be monitored.
$k_\text{max}$ is selected so that the prior probability of occurrence of more than $k_\text{max}$ simultaneous faults is much smaller than the integrity risk budget.
This probability is denoted as $P_{\text{not monitored}}$.
The procedure for determining $k_\text{max}$ and $P_{H_i}$ is detailed in \citet{blanch2015baseline} and will not be elaborated on here.

\subsection{Reconstruction of Jackknife Residual}\label{subsec:reconstruct_JK}
For fault mode $k$, the weight matrix in Eq.
\eqref{eq:S^i} can be re-constructed as follows:
\begin{equation}
    W^{(k)}_{i,i} = \begin{cases}
         0 & \text{if}~ i \in idx_k^{ex} \\
         W_{i,i} & \text{otherwise}
\end{cases}\,.
\end{equation}
The jackknife residual regarding the $i \in idx_k^{ex}$th measurement for fault mode $k$ is given by 
\begin{equation}
    t_i^{(k)} = y_i - \hat{y}_i^{(k)} \,,
\end{equation}
where $\hat{y}_i^{(k)}$ is the predicted $i$th measurement based on subset solution $\hat{\mathbf{x}}^{(k)}$, as defined in Eq.
\eqref{eq:prediction}.
It is easy to extend Eq.
\eqref{eq:ti_3} to the simultaneous faults condition as follows:
\begin{equation}
    t_i^{(k)} = \sum_{j=1}^n \tilde{p}^{(k)}_{i,j}\varepsilon_j 
\label{eq:define_statistic}, i \in idx_k^{ex} \,,
\end{equation}
where  $\tilde{p}^{(k)}_{i,j}$ is the $(i,j)$ element of $\mathbf{I}-\tilde{\mathbf{P}}^{(k)}$.
\par
It is worth noting that the existence of $t_i^{(k)}$ depends on the existence of the subset solution $\hat{\mathbf{x}}^{(k)}$, which is not guaranteed in the constellation fault mode.
This is because all satellite measurements from the hypothetically faulty constellation are excluded in this fault mode, making it impossible to solve the receiver clock bias related to the hypothetically faulty constellation in  $\hat{\mathbf{x}}^{(k)}$.
Therefore, the constellation fault is temporarily not considered in constructing jackknife detectors in the following sections.
This problem will be reviewed in Section \ref{sec:consider_constellation}.

\subsection{Combination of Jackknife Residuals}
\begin{review}
When $k>n$, there are multiple jackknife residuals associated with fault mode $k$, which requires a vector-form monitoring strategy to be statistically consistent with multiple-hypothesis solution separation (MHSS). Specifically, the following triple-axis combination of jackknife residuals is adopted:
\end{review}
\begin{equation}
    \tilde{t}_{k,v} = \sum_{i \in idx_k^{ex}} S_{v,i}t_i^{(k)}, ~~k=n+1,n+2,\cdots, N_\text{fault modes},~~v=1,2,3 \,,
    \label{eq:combint_JK}
\end{equation}
where $S_{v,i}$ is the $(v,i)$th element of the full set solution matrix $\mathbf{S}$. This kind of weighting scheme can greatly reduce the complexity of developing integrity monitoring algorithms, as will be shown in Section \ref{sec:IR_eva}. \begin{review}Moreover, the jacknife detector and the solution separation detector are equivalent under this weighting scheme, as been proved in \ref{sec:app_SS_JK}\end{review}.\par
By substituting Eq.
\eqref{eq:define_statistic} into Eq.
\eqref{eq:combint_JK}, we have
\begin{equation}
\tilde{t}_{k,v} = \sum_{j=1}^n \sum_{i \in idx_k^{ex}} S_{v,i}\tilde{p}^{(k)}_{i,j}\varepsilon_j \,.
\label{eq:ti_tilde}
\end{equation}
The PDF of $\tilde{t}_{k,v}$ can be derived as
\begin{equation}
\begin{aligned}
    f_{\tilde{t}_{k,v}}(x) = {\prod\limits_{j = 1}^{n}\left| \sum_{i \in idx_k^{ex}}S_{v,i}\tilde{p}_{i,j}^{(k)} \right|^{- 1}}  &f_{\varepsilon_1}\left( \frac{x}{\left| \sum_{i \in idx_k^{ex}}S_{v,i}\tilde{p}_{i,1}^{(k)} \right|} \right)*f_{\varepsilon_2}\left( \frac{x}{\left| \sum_{i \in idx_k^{ex}}S_{v,i}\tilde{p}_{i,2}^{(k)} \right|} \right)* \\
&\ldots*f_{\varepsilon_n}\left( \frac{x}{\left| \sum_{i \in idx_k^{ex}}S_{v,i}\tilde{p}_{i,n}^{(k)} \right|} \right) \,.
    \label{eq:ti_conv_multifault} 
\end{aligned}
\end{equation}
In the special case of Gaussian errors, i.e., $\varepsilon_j \sim \mathcal{N}(0,\sigma_j^2)$, we have
\begin{equation}
\tilde{t}_{k,v} \sim \mathbb{N}\left(0,\sum_{j=1}^n \left(\sum_{i \in idx_k^{ex}} S_{v,i}\tilde{p}^{(k)}_{i,j}\right)^2 \sigma_j^2\right) \,.
    \label{eq:ti_dis_multifault}
\end{equation}
\begin{review}
In this work, we adopt a hybrid monitoring logic:
\begin{itemize}
    \item For single-fault modes ($k\leq n$), we use a single scalar jackknife residual $t_k$;
    \item For multi-fault modes ($k>n$), we use tripke-axis test statistics $\tilde{t}_{k,1}$ (East), $\tilde{t}_{k,2}$ (North), and $\tilde{t}_{k,3}$ (Up).
\end{itemize}
\end{review}

\subsection{Reconstruction of Hypothesis Tests}
The following hypotheses are constructed:
\begin{review}
\begin{equation}
\begin{aligned}
    H_0{:}& ~\text{The hypothesis corresponding to fault mode} ~ 0 \\
    H_k{:}& ~\text{The hypothesis corresponding to fault mode} ~ k \,,
\end{aligned}  
\nonumber
\end{equation}
which involves a multiple testing problem.
The reject regions for test $H_0$ v.s.
$H_k$ can be defined as
\begin{equation}
    \begin{cases}
        R_{k} = \{t_{k} | ~|t_{k}| \geq T_{k}\} & k=1, 2, \cdots, n \\
        R_{k,v}=\{\tilde{t}_{k,v} | ~|\tilde{t}_{k,v}| \geq T_{k,v}\} & k=n+1, n+2, \cdots, N_\text{fault modes},~v=1,2,3
    \end{cases} \,,
\end{equation}
where $T_k$ and $T_{k,v}$ are the thresholds for the scalar and triple-axis monitors, respectively.
Assume that the probability of the Type I error of the above multiple testing problem is $\alpha^*$, i.e.,
\begin{equation}
\alpha^* = P\Big(\bigcup_{k=1}^{n} t_{k} \in R_{k} ~\cup~ \bigcup_{k=n+1}^{N_\text{fault modes}} \bigcup_{v=1}^{3} \tilde{t}_{k,v} \in R_{k,v} ~\Big|~ H_0 \Big) \,.
\end{equation}
Since the reject regions are not mutually exclusive, we have
\begin{equation}
\begin{aligned}
\alpha^* \leq&~ \sum_{k=1}^{n} P\Big( |t_{k}| \geq T_{k} ~\Big|~ H_0\Big) + \sum_{k=n+1}^{N_\text{fault modes}}\sum_{v=1}^{3} P\Big( |\tilde{t}_{k,v}| \geq T_{k,v} ~\Big|~ H_0\Big) =\tau \,.
\end{aligned}
    \label{eq:ineq_mul}
\end{equation}
According to the Bonferroni correction \citep{bonferroni1936teoria}, by setting 
\begin{equation}
        T_{k,v} = \begin{cases}
            Q^{-1}_{\tilde{t}_k}\left(\frac{\tau}{2 N_\text{fault modes}}\right) & k\leq n\\
            Q^{-1}_{\tilde{t}_{k,v}}\left(\frac{\tau_v}{2 N_\text{fault modes}}\right) & k>n
        \end{cases}
        \label{eq:threshold_jk_multi} \,,
\end{equation}
$H_0$ is rejected if any active monitor enters the reject region at significant level of $\alpha^*$, where $\sum_{v=1,2,3}\tau_v=\tau$, $Q^{-1}_{\tilde{t}_{k,v}}(\cdot)$ is the quantile function of the distribution of $\tilde{t}_{k,v}$, and $\tau$ is a user-defined value.
Eq.~\eqref{eq:ineq_mul} indicates that $\tau$ is the upper limit of $\alpha^*$. It is worth noting that the Bonferroni correction is overly conservative, which can raise miss-detection risks. This is a common issue in multiple hypothesis testing, which also exists in the \textcolor{black}{SS ARAIM} algorithm. A detailed analysis of the Bonferroni correction is provided in Section II.3 and Appendix C of \citet{yan2025efficient}.
\end{review}

\section{Jackknife RAIM with non-Gaussian Nominal Errors}\label{sec:JackRAIM}
This section develops a multiple-hypothesis-based integrity monitoring algorithm based on the improved jackknife detector in Section \ref{sec:multiple_JK}, aiming to deal with non-Gaussian nominal error bounds.
The proposed method is named the jackknife ARAIM algorithm to emphasize its usage of the jackknife detector.
The jackknife ARAIM algorithm follows a similar process to the \textcolor{black}{SS ARAIM} algorithm, beginning with defining the threat model, constructing the fault detectors, and determining their threshold to comply with the continuity requirements, then evaluating integrity risks, and concluding with deriving protection levels.
The principal difference between the proposed jackknife ARAIM algorithm and the \textcolor{black}{SS ARAIM} algorithm lies in the choice of fault detectors.
Instead of using solution separation in the position domain, the proposed method systematically exploits the properties of the jackknife detector in the range domain and derives a tight bound of the integrity risk.

\subsection{Determine the Threshold of Monitors}\label{sec:threshold_monitors}
The threshold of monitors, i.e., jackknife detectors, is determined so that the continuity requirement is satisfied.
Given the continuity budget caused by false alerts $C_{\text{REQ,FA}}$, the continuity risk can be written as follows:
\begin{review}
\begin{equation}
    P_{\text{continuity}} = P\Big(\bigcup_{k=1}^{n} t_{k} \in R_{k} ~\cup~ \bigcup_{k=n+1}^{N_{\text{fault modes}}}\bigcup_{v=1}^{3} \tilde{t}_{k,v} \in R_{k,v} ~\Big|~ H_0 \Big) P_{H_0} \leq C_\text{REQ,FA}\,,
    \label{eq:continuity define JK}
\end{equation}
Since the reject regions are not mutually exclusive, we have
\begin{equation}
\begin{aligned}
P_{\text{continuity}} \leq&~ \sum_{k=1}^{n} P\Big( |t_{k}| \geq T_{k} ~\Big|~ H_0\Big)P_{H_0} + \sum_{k=n+1}^{N_\text{fault modes}} \sum_{v=1}^{3} P\Big( |\tilde{t}_{k,v}| \geq T_{k,v} ~\Big|~ H_0\Big)P_{H_0} \,.
\end{aligned}
\end{equation}
The threshold $T_{k,v}$ is determined by the allocated continuity budget caused by false alert as follows:
\begin{equation}
  T_{k,v} = \begin{cases}
     Q^{-1}_{\tilde{t}_{k,v}}\left(\frac{C_\text{REQ,FA}}{2N_\text{fault modes}P_{H_0}}\right) & k\leq n\\
     Q^{-1}_{\tilde{t}_{k,v}}\left(\frac{C_\text{REQ,FA,v}}{2 N_\text{fault modes}P_{H_0}}\right) & k>n
  \end{cases} \,,
    \label{eq:threshold_continuity}
\end{equation}
where $C_\text{REQ,FA,v}$ is the allocated continuity budget caused by a false alert in the $v$th direction\end{review}. An equal allocation strategy is adopted in allocating the continuity budget to each fault mode, which is the same as that in the \textcolor{black}{SS ARAIM} algorithm.

As shown in Eqs.
\eqref{eq:ti_3} and \eqref{eq:ti_tilde}, $t_k$ and $\tilde{t}_{k,v}$ are the linear combination of nominal measurement error bounds, i.e., $\varepsilon_1, \varepsilon_2, \cdots, \varepsilon_n$.
Here, $\varepsilon_j,j=1,2\cdots,n$ refers to the nominal error bound for accuracy.
The quantile function $Q^{-1}_{t_k}(\cdot)$ and $Q^{-1}_{\tilde{t}_{k,v}}(\cdot)$ can be evaluated by using the numerical method developed in \citet{yan2024principal}.\par

\subsection{Integrity Risk Evaluation}\label{sec:IR_eva}
The detection threshold determined in Eq.
\eqref{eq:threshold_continuity} can be used to evaluate the integrity risk as follows:
\begin{review}
\begin{equation}
    P_{\text{HMI}} = \sum_{i=0}^{N_{\text{fault modes}}} P\Big(\{|e_v|>\ell_v \}\cap \bigcap_{k=1}^{n} \{|t_k| < T_k\} \cap \bigcap_{k=n+1}^{N_{\text{fault modes}}}\bigcap_{v=1}^{3} \{|\tilde{t}_{k,v}| < T_{k,v}\} ~\Big|~ H_i\Big) P_{H_i} + P_{\text{not monitored}} \leq I_{\text{REQ}} \,,
    \label{eq:Int_req}
\end{equation}
where $e_v$ is the positioning error component in the $v$th direction
\begin{equation}
    e_v = (\hat{\mathbf{x}} - \mathbf{x})_v \,,
    \label{eq:error_of_interest}
\end{equation}
\end{review}
and $\ell_v$ is the alert limit in the $v$th direction; and $I_\text{REQ}$ is the integrity budget.
\par
Let 
\begin{review}
\begin{equation}
    I_{cal} = \sum_{i=0}^{N_{\text{fault modes}}} P\Big(\{|(\hat{\mathbf{x}} - \mathbf{x})_v|>\ell_v \} \cap \bigcap_{k=1}^{n} \{|t_k| < T_k\} \cap \bigcap_{k=n+1}^{N_{\text{fault modes}}}\bigcap_{v=1}^{3} \{|\tilde{t}_{k,v}| < T_{k,v}\} ~\Big|~ H_i\Big) P_{H_i} \,,
    \label{eq:integrity_risk_cal}
\end{equation}
\end{review}
which is the sum of hazardously misleading information (HMI) probabilities over the fault-free hypothesis and other faulted hypotheses.
\par
\subsubsection{Bound on the probability of HMI under $H_0$}
In the fault-free hypothesis $H_0$, a bound on the probability of HMI is established as follows
\begin{review}
\begin{equation}
P\Big(\{|(\hat{\mathbf{x}} - \mathbf{x})_v|>\ell_v\} \cap \bigcap_{k=1}^{n} \{|t_k| < T_k\} \cap \bigcap_{k=n+1}^{N_{\text{fault modes}}}\bigcap_{v=1}^{3} \{|\tilde{t}_{k,v}| < T_{k,v}\} ~\Big|~ H_0\Big) \leq  P(|(\hat{\mathbf{x}} - \mathbf{x})_v|>\ell_v|H_0)\,.
    \label{eq:HMI_bound_H0}
\end{equation}
\end{review}
This bound is obtained by ignoring knowledge of no detection, which can be considered a tight bound \citep{joerger_solution_2014}.
This is because the probability of no detection under the fault-free hypothesis is larger than $1-C_\text{REQ,FA}$, as ensured by Eq.
\eqref{eq:continuity define JK}.\par
By substituting Eqs.
\eqref{eq:general_linear} and \eqref{eq:full solution} into $(\hat{\mathbf{x}} - \mathbf{x})_v$, we have
\begin{equation}
    (\hat{\mathbf{x}} - \mathbf{x})_v = (\mathbf{S}\bm{\varepsilon})_v = \sum_{i=1}^{n} S_{v,i} \varepsilon_i \,,
\end{equation}
where $S_{v,i}$ is the $(v,i)$th element in $\mathbf{S}$.
Then the PDF of $(\hat{\mathbf{x}} - \mathbf{x})_v$ is given by
\begin{equation}
f_{(\hat{\mathbf{x}} - \mathbf{x})_v}(t) = {\prod\limits_{j = 1}^{n}\left| S_{v,i} \right|^{- 1}} f_{\varepsilon_1}\left( \frac{t}{\left|S_{v,1} \right|} \right)*f_{\varepsilon_2}\left( \frac{t}{\left| S_{v,2} \right|} \right)*\ldots  * f_{\varepsilon_n}\left( \frac{t}{\left| S_{v,n} \right|} \right) \,.
    \label{eq:pdf_position_error}
\end{equation}
Eq.
\eqref{eq:pdf_position_error} can be used to evaluate the bound in Eq.
\eqref{eq:HMI_bound_H0}.\par

\subsubsection{Bound on the probability of HMI under $H_k$}
In the faulted hypothesis $H_k$, a similar bound on the probability of HMI is given as follows:
\begin{review}
\begin{equation}
\begin{aligned}
&P\Big(\{|(\hat{\mathbf{x}} - \mathbf{x})_v|>\ell_v\}\cap \bigcap_{j=1}^{n} \{|t_j| < T_j\} \cap \bigcap_{j=n+1}^{N_{\text{fault modes}}}\bigcap_{u=1}^{3} \{|\tilde{t}_{j,u}| < T_{j,u}\} ~\Big|~ H_k\Big) \\
&\leq  P\Big(\{|(\hat{\mathbf{x}} - \mathbf{x})_v|>\ell_v\}\cap \mathcal{A}_k ~\Big|~H_k\Big)\,,
\end{aligned}
    \label{eq:ignore_no_detect_Hk}
\end{equation}
where the active no-alert event under fault mode $k$ is
\begin{equation}
\mathcal{A}_k =
\begin{cases}
\{|t_k| < T_k\} & k\leq n\\
\bigcap_{u=1}^{3} \{|\tilde{t}_{k,u}| < T_{k,u}\} & k>n
\end{cases}\,.
\label{eq:active_no_alert}
\end{equation}
\end{review}
Again, this bound is obtained by ignoring knowledge of no detection for all other hypothesis tests, except for the one for the test $H_0$ v.s.
$H_k$.
As proven in \citet{joerger_solution_2014}, Eq.
\eqref{eq:ignore_no_detect_Hk} also provides a tight bound on the probability of HMI under $H_k$.
The right-hand side of Eq.
\eqref{eq:ignore_no_detect_Hk} can be simplified by invoking the conditional probability
\begin{review}
\begin{equation}
\begin{aligned}
    P(\{|(\hat{\mathbf{x}} - \mathbf{x})_v|>\ell_v\}\cap \mathcal{A}_k |H_k) & =  P(|(\hat{\mathbf{x}} - \mathbf{x})_v|>\ell_v| H_k \cap \mathcal{A}_k) P(\mathcal{A}_k | H_k) \\
&\leq P(|(\hat{\mathbf{x}} - \mathbf{x})_v|>\ell_v| H_k \cap \mathcal{A}_k) \,.
    \end{aligned}
    \label{eq:HMI_bound_HK}
\end{equation}
The inequality in the second line bounds $P(\mathcal{A}_k | H_k)$ with $P(\mathcal{A}_k | H_k)=1$.
\par
\end{review}
A further relaxation of Eq.
\eqref{eq:HMI_bound_HK} is achieved by exploiting the structure of $ (\hat{\mathbf{x}} - \mathbf{x})_v$ under $H_k$.
Define the fault vector in the faulted hypothesis $H_k$ as $\mathbf{b}^{(k)}$.
This $n\times 1$ vector  takes the following form:
\begin{equation}
    b^{(k)}_j =\begin{cases}
        b_j & \text{if}~j\in idx_k^{ex} \\
        0 & \text{otherwise}
    \end{cases} \,,
\end{equation}
where $ b^{(k)}_j$ is the $j$th element of $\mathbf{b}^{(k)}$ and $b_j,j=1,2,\cdots, n$ is an unknown constant with non-zero values.
In the faulted hypothesis $H_k$, the linearized measurement model can be written as
\begin{equation}
    \mathbf{y} = \mathbf{G}\mathbf{x}+\bm{\varepsilon}+\mathbf{b}^{(k)} \,,
    \label{eq:meas_model_bias}
\end{equation}
where
\begin{equation}
    y_j = \begin{cases}
        \mathbf{g}_j \mathbf{x}+ \varepsilon_j + b_j & \text{if}~ j \in idx_k^{ex}\\
        \mathbf{g}_j \mathbf{x} + \varepsilon_j & \text{otherwise} \,,
    \end{cases}
    \label{eq:meas_model_bias2}
\end{equation}
and $\mathbf{g}_j$ is the $j$th row of $\mathbf{G}$.
\par

Different from Section \ref{sec:threshold_monitors}, $\varepsilon_i, i=1,2,\cdots,n$ in Eqs.
\eqref{eq:meas_model_bias} and \eqref{eq:meas_model_bias2} refers to the nominal measurement error bound for integrity.
This kind of bound considers the effects of nominal signal deformation errors, which is realized by introducing a $b_{nom}$ term to create two equally shifted nominal measurement error bounds for accuracy.
To simplify the derivation, we first ignore the effects of nominal signal deformation errors by setting $b_{nom,i}=0, i=1,2,\cdots,n$.
Then the nominal measurement error bound for integrity is the same as that for accuracy.
\par

Now, $ (\hat{\mathbf{x}} - \mathbf{x})_v$ under $H_k$ can be written by
\begin{equation}
\begin{aligned}
    (\hat{\mathbf{x}} - \mathbf{x})_v |H_k & =(\mathbf{S}\mathbf{y}-\mathbf{x})_v | H_k\\
    & = \left(\mathbf{S}(\mathbf{G}\mathbf{x}+\bm{\varepsilon}+\mathbf{b}^{(k)}) - \mathbf{x} \right)_v\\
    & = \left(\mathbf{S}\bm{\varepsilon}+ \mathbf{S}\mathbf{b}^{(k)} \right)_v \\
& = \sum_{i=1}^{n}S_{v,i}\varepsilon_i + \sum_{j\in idx_k^{ex}} S_{v,j}b_j \,.
\end{aligned}
\label{eq:error_of_interest_3}
\end{equation}
For each $j\in idx_k^{ex}$, the corresponding jackknife residual is given by
\begin{equation}
\begin{aligned}
    t_j^{(k)} &=  y_j - \hat{y}_j \\
    & =\mathbf{g}_j \mathbf{x}+\varepsilon_j + b_j - \mathbf{g}_j \hat{\mathbf{x}}^{(k)} \\
    & = \mathbf{g}_j (\mathbf{x} - \hat{\mathbf{x}}^{(k)})+\varepsilon_j + b_j  \\
& = -\mathbf{g}_j \mathbf{S}^{(k)}\bm{\varepsilon}+\varepsilon_j + b_j  \,.
\end{aligned}
\label{eq:jack_res_find}
\end{equation}
The last line holds because $ \mathbf{x} - \hat{\mathbf{x}}^{(k)} = \mathbf{S}^{(k)}\bm{\varepsilon}$.
Then, we have
\begin{equation}
b_j =t_j^{(k)} + \mathbf{g}_j\mathbf{S}^{(k)}\bm{\varepsilon} - \varepsilon_j \,.
    \label{eq:bias}
\end{equation}
By substituting Eq.
\eqref{eq:bias} into Eq.
\eqref{eq:error_of_interest_3}, we have
\begin{equation}
\begin{aligned}
    (\hat{\mathbf{x}} - \mathbf{x})_v | H_k&= \sum_{i=1}^{n}S_{v,i}\varepsilon_i + \sum_{j\in idx_k^{ex}} S_{v,j}(t_j^{(k)} + \mathbf{g}_j\mathbf{S}^{(k)}\bm{\varepsilon} - \varepsilon_j)\\
&=\sum_{j \not \in idx_k^{ex}} S_{v,j}\varepsilon_j +\sum_{j\in idx_k^{ex}} S_{v,j}\mathbf{g}_j\mathbf{S}^{(k)} \bm{\varepsilon}+\sum_{j\in idx_k^{ex}} S_{v,j}t_j^{(k)} \,.
\end{aligned}
\label{eq:error_of_interest_4}
\end{equation}
Let $\mathbf{E}^{(k)}$ be a $n\times n$ diagonal matrix with the following definition
\begin{equation}
    E^{(k)}_{j,j} = \begin{cases}
        0 & \text{if}~ j \in idx_k^{ex}\\
        1 & \text{otherwise}
\end{cases} \,.
\end{equation}
Eq.
\eqref{eq:error_of_interest_4} can be simplified to
\begin{equation}
    (\hat{\mathbf{x}} - \mathbf{x})_v |H_k= \mathbf{q}^{(k)} \bm{\varepsilon}+ \sum_{j\in idx_k^{ex}} S_{v,j}t_j^{(k)} \,,
\end{equation}
where
\begin{equation}
\mathbf{q}^{(k)} = \mathbf{s}_v \mathbf{E}^{(k)} + \sum_{j\in idx_k^{ex}}  S_{v,j}\mathbf{g}_j\mathbf{S}^{(k)} \,.
   \label{eq:qk}
\end{equation}
The distribution of $\mathbf{q}^{(k)} \bm{\varepsilon} $ is given by 
\begin{equation}
    f_{\mathbf{q}^{(k)} \bm{\varepsilon}}(x) = {\prod\limits_{j = 1}^{n}\left| q^{(k)}_{j} \right|^{- 1}} f_{\varepsilon_1}\left( \frac{x}{\left| q^{(k)}_{1} \right|} \right)*f_{\varepsilon_2}\left( \frac{x}{\left| q^{(k)}_{2} \right|} \right)*\ldots  * f_{\varepsilon_n}\left( \frac{x}{\left| q^{(k)}_{n} \right|} \right) \,,
    \label{eq:pdf_qk}
\end{equation}
where $q^{(k)}_j,j=1,2,\cdots,n$ is the $j$th element of $\mathbf{q}^{(k)}$.
\par
Then the bound on the probability of HMI under $H_k$ in Eq.
\eqref{eq:HMI_bound_HK} can be written by
\begin{review}
\begin{equation}
\begin{aligned}
    P(|(\hat{\mathbf{x}} - \mathbf{x})_v |>\ell_v| H_k \cap \mathcal{A}_k ) &= P(|\mathbf{q}^{(k)} \bm{\varepsilon}+ \sum_{j\in idx_k^{ex}} S_{v,j}t_j^{(k)}|>\ell_v | H_k \cap \mathcal{A}_k)\\
&\leq P(|\mathbf{q}^{(k)} \bm{\varepsilon}|+ |\sum_{j\in idx_k^{ex}} S_{v,j}t_j^{(k)}|>\ell_v | H_k\cap \mathcal{A}_k) \,.
\end{aligned}
\label{eq:HMI_bound_HK2}
\end{equation}
The second line holds because of the triangle inequality.
\par
When $k\leq n$, $\mathcal{A}_k=\{|t_k|<T_k\}$.
Then, the right-hand side of Eq.
\eqref{eq:HMI_bound_HK2} can be written by
\begin{equation}
P(|\mathbf{q}^{(k)}\bm{\varepsilon}|+|S_{v,k}t_k|>\ell_v | H_k \cap \{t_k \leq T_k\}) \leq P(|\mathbf{q}^{(k)}\bm{\varepsilon}|+|S_{v,k}|T_k>\ell_v | H_k ) \,.
    \label{eq:HMI_bound_HK3_type1}
\end{equation}
When $k>n$, $\mathcal{A}_k=\bigcap_{u=1}^{3} \{|\tilde{t}_{k,u}|<T_{k,u}\}$ and $\tilde{t}_{k,v} = \sum_{j\in idx_k^{ex}} S_{v,j}t_j^{(k)}$.
Then, the right-hand side of Eq.
\eqref{eq:HMI_bound_HK2} can be written by
\begin{equation}
\begin{aligned}
    &P(|\mathbf{q}^{(k)} \bm{\varepsilon}|+ |\tilde{t}_{k,v}|>\ell_v ~\Big|~ H_k\cap \bigcap_{u=1}^{3} \{|\tilde{t}_{k,u}| < T_{k,u}\}) \\
    &\leq P(|\mathbf{q}^{(k)} \bm{\varepsilon}|+ |\tilde{t}_{k,v}|>\ell_v ~\Big|~ H_k\cap |\tilde{t}_{k,v}| < T_{k,v}) \\
&\leq P(|\mathbf{q}^{(k)}\bm{\varepsilon}|+T_{k,v}>\ell_v | H_k ) \,.
\end{aligned}
\label{eq:HMI_bound_HK3_type2}
\end{equation}
\end{review}
\subsubsection{Finalized bound of integrity risk}\label{subsec:final_int_risk}
Finally, the bound of integrity risk for monitored fault modes in Eq.
\eqref{eq:integrity_risk_cal} is given by summarizing Eqs.
\eqref{eq:HMI_bound_H0}, \eqref{eq:HMI_bound_HK3_type1} and \eqref{eq:HMI_bound_HK3_type2} as follows:
\begin{review}
\begin{equation}
\begin{aligned}
    I_{cal} &\leq  P(|(\hat{\mathbf{x}} - \mathbf{x})_v|>\ell_v|H_0) P_{H_0} \\
    &+ \sum_{k=1}^{n} P(|\mathbf{q}^{(k)}\bm{\varepsilon}|+|S_{v,k}|T_k>\ell_v | H_k )P_{H_k}  \\
   & + \sum_{k=n+1}^{N_\text{fault modes}}P(|\mathbf{q}^{(k)}\bm{\varepsilon}|+T_{k,v}>\ell_v | H_k) P_{H_k} \\
  & \leq I_\text{REQ}^v\left(1- \frac{P_\text{not monitored}}{I_\text{REQ}} \right) \,,
\end{aligned}
\label{eq:I_cal_bound}
\end{equation}
\end{review}
where $I_\text{REQ}^3$ standards for the integrity budget for the vertical component, $I_\text{REQ}^1+I_\text{REQ}^2$ represents the integrity budget for the horizontal component, and $I_\text{REQ}^1=I_\text{REQ}^2$.
Notably, the distributions of $(\hat{\mathbf{x}} - \mathbf{x})_v$ and $\mathbf{q}^{(k)}\bm{\varepsilon}$ are known, as given in Eqs.
\eqref{eq:pdf_position_error} and \eqref{eq:pdf_qk}, respectively.
Hence, the inequality condition in the last line can be evaluated to check if the integrity requirement is satisfied.\par
So far, we have derived the bound of integrity risk for monitored fault modes with $b_{nom}=0$.
To consider the effects of nominal signal deformation errors, Eq.
\eqref{eq:I_cal_bound} can be modified as follows:
\begin{review}
\begin{equation}
\begin{aligned}
    I_{cal} &\leq  P(|(\hat{\mathbf{x}} - \mathbf{x})_v|>\ell_v - b_v^{(0)}|H_0) P_{H_0} \\
    &+ \sum_{k=1}^{n} P(|\mathbf{q}^{(k)}\bm{\varepsilon}|+|S_{v,k}|T_k>\ell_v - b_v^{(k)}| H_k )P_{H_k}  \\
   & + \sum_{k=n+1}^{N_\text{fault modes}}P(|\mathbf{q}^{(k)}\bm{\varepsilon}|+T_{k,v}>\ell_v - b_v^{(k)}| H_k) P_{H_k} \\
  & \leq I_\text{REQ}^v\left(1- \frac{P_\text{not monitored}}{I_\text{REQ}} \right) \,,
\end{aligned}
\label{eq:I_cal_bound_final}
\end{equation}
\end{review}
where $b_v^{(k)}$ represents the worst-case impact of nominal signal deformation errors on the position solution:
\begin{equation}
b_v^{(k)} = \sum_{i=1}^{n}|S_{v,i}^{(k)}|b_{nom,i} \,.
    \label{eq:bv}
\end{equation}
\subsection{Protection Level Derivation}\label{sec:PL}
By replacing the alert limit $\ell_v$ with the protection level $PL_v$ and replacing the last inequality with equality in Eq.
\eqref{eq:I_cal_bound_final}, the PL can be derived as follows:
\begin{review}
\begin{equation}
\begin{aligned}
    &P(|(\hat{\mathbf{x}} - \mathbf{x})_v|>PL_v-b_v^{(0)}|H_0) P_{H_0} \\
    &+ \sum_{k=1}^{n} P(|\mathbf{q}^{(k)}\bm{\varepsilon}|+|S_{v,k}|T_k>PL_v-b_v^{(k)} | H_k )P_{H_k}  \\
   & + \sum_{k=n+1}^{N_\text{fault modes}}P(|\mathbf{q}^{(k)}\bm{\varepsilon}|+T_{k,v}>PL_v -b_v^{(k)}| H_k) P_{H_k} \\
& =  I_\text{REQ}^v\left(1- \frac{P_\text{not monitored}}{I_\text{REQ}} \right) \,.
\end{aligned}
\label{eq:I_cal_bound2}
\end{equation}
\end{review}
To solve $PL_v$, the integrity budget $I_\text{REQ}^v\left(1- \frac{P_\text{not monitored}}{I_\text{REQ}} \right)$ needs to be allocated to each fault mode.
Specifically, $PL_v$ is given by
\begin{review}
\begin{equation}
\begin{aligned}
    PL_v = \max \Biggl \{  & Q^{-1}_{(\hat{\mathbf{x}} - \mathbf{x})_v} \left(\frac{I_{REQ,0}^v}{2P_{H_0}}\right) + b_v^{(0)},  
    \max_{1<k\leq n} \left\{  Q^{-1}_{\mathbf{q}^{(k)}\bm{\varepsilon}} \left(\frac{I_{REQ,k}^v}{2P_{H_k}}\right) + |S_{v,k}|T_k + b_v^{(k)}\right\} ,\\
    & \max_{n<k\leq N_\text{fault modes}} \left\{  Q^{-1}_{\mathbf{q}^{(k)}\bm{\varepsilon}} \left(\frac{I_{REQ,k}^v}{2P_{H_k}}\right) + T_{k,v} + b_v^{(k)}\right\} \Biggl\} \,,
\end{aligned}
\label{eq:PL_final}
\end{equation}
\end{review}
where
\begin{equation}
\sum_{k=1}^{N_\text{fault modes}} I_{REQ,k}^v = I_\text{REQ}^v\left(1- \frac{P_\text{not monitored}}{I_\text{REQ}} \right) \,.
\end{equation}
The quantile functions $Q^{-1}_{(\hat{\mathbf{x}}-\mathbf{x})_v}$ and $Q^{-1}_{\mathbf{q}^{(k)}\bm{\varepsilon}}$ can be evaluated by using the numerical method developed in \citet{yan2024principal}.
\par
In this paper, the equal allocation strategy for integrity is applied as follows:
\begin{equation}
I_{REQ,k}^v  = \frac{1}{N_\text{fault modes}} I_\text{REQ}^v\left(1- \frac{P_\text{not monitored}}{I_\text{REQ}} \right) \,.
     \label{eq:equal_allocation_I_REQ}
\end{equation}
\begin{review}Note that it is applicable to apply optimal allocation strategies for allocating continuity and integrity risk budgets in Eq. \eqref{eq:threshold_continuity} and Eq. \eqref{eq:equal_allocation_I_REQ}, respectively. The least friction way is to use the allocation strategy described in \citet{blanch2015baseline} as the proposed method shares a similar logic with the \textcolor{black}{SS ARAIM}. Other optimization-based allocation strategies can also be applied, such as the improved Cuckoo search algorithm \citep{yan2025araim}, which incorporates a subset grouping method with a feedback structure to reduce the number of fault subsets, and the sparrow search algorithm \citep{mu2024araim}, which optimizes risk allocation and position solution simultaneously. These methods are designed for the \textcolor{black}{SS ARAIM} algorithm, and therefore they can also be applied to the proposed jackknife ARAIM algorithm with minor modifications.
\end{review}

The vertical protection level (VPL) is directly given by $PL_3$, i.e.,
\begin{equation}
    VPL = PL_3 \,,
\end{equation}
and the horizontal protection level (HPL) is given by synthesizing $PL_1$ and $PL_2$ as follows:
\begin{equation}
HPL = \sqrt{PL_1^2+PL_2^2} \,.
\end{equation}

\subsection{Consideration of Constellation Faults}\label{sec:consider_constellation}
As discussed in Section \ref{sec:multiple_JK}, the jackknife residual is not computable in the constellation fault mode.
Therefore, the PL calculation in Section \ref{sec:PL} does not consider constellation fault modes.
However, it is essential to consider the possibility of constellation faults in the multi-constellation system to protect integrity.
To address this issue, one can use the solution separation detector to construct the hypothesis regarding the constellation fault and integrate it into the PL equations in Section \ref{sec:PL}.
\par
Let  $\Omega_\text{const}$ be the set of fault modes involving constellation faults.
Under each fault mode $k \in \Omega_\text{const}$, the integrity risk of HMI is given by
\begin{equation}
    P(\{|(\hat{\mathbf{x}} - \mathbf{x})_v|>\ell_v\}\cap \{|d_v^{(k)}| < D_{k,v}\} | H_k, k \in \Omega_\text{const}) \,,
    \label{eq:HMI_const}
\end{equation}
where $d_v^{(k)} = (\hat{\mathbf{x}} - \hat{\mathbf{x}}^{(k)})_v$ and $D_{k,v}$ are the solution separation test statistic and its threshold, respectively \citep{blanch2015baseline}.
According to the triangular inequality,
\begin{equation}
|(\hat{\mathbf{x}} - \mathbf{x})_v| = |(\hat{\mathbf{x}} - \hat{\mathbf{x}}^{(k)} + \hat{\mathbf{x}}^{(k)} - \mathbf{x})_v| \leq |(\hat{\mathbf{x}} - \hat{\mathbf{x}}^{(k)})_v| + | (\hat{\mathbf{x}}^{(k)} - \mathbf{x})|_v \,.
\end{equation}
Therefore, Eq.
\eqref{eq:HMI_const} can be bounded by
\begin{subequations}
\begin{align}
    & P(\{|(\hat{\mathbf{x}} - \mathbf{x})_v|>\ell_v\}\cap \{|d_v^{(k)}| < D_{k,v} \}| H_k, k \in \Omega_\text{const}) \\
    & \leq P(\{|(\hat{\mathbf{x}} - \hat{\mathbf{x}}^{(k)})_v| + | (\hat{\mathbf{x}}^{(k)} - \mathbf{x})_v| >\ell_v\}\cap \{|d_v^{(k)}| < D_{k,v}\} | H_k, k \in \Omega_\text{const}) \\
    & \leq  P(| (\hat{\mathbf{x}}^{(k)} - \mathbf{x})_v| + D_{k,v} >\ell_v | H_k, k \in \Omega_\text{const}) \,,
\end{align}
\end{subequations}
\begin{review}where $(\hat{\mathbf{x}}^{(k)} - \mathbf{x})_v=\sum_{j=1}^{n}\tilde{s}_{v,j}\varepsilon_j$ is the linear combination of nominal measurement erorr bounds \citep{yan2025efficient}, and $\tilde{s}_{v,j}$ is the is the element in the $v$th row and $j$th column of $\mathbf{S}-\mathbf{S}^{(k)}$\end{review}.
Following the steps in Section \ref{subsec:final_int_risk}, Eq.
\eqref{eq:I_cal_bound2} can be eventually re-written as
\begin{review}
\begin{equation}
\begin{aligned}
    &P(|(\hat{\mathbf{x}} - \mathbf{x})_v|>PL_v-b_v^{(0)}|H_0) P_{H_0} \\
    &+ \sum_{k=1}^{n} P(|\mathbf{q}^{(k)}\bm{\varepsilon}|+|S_{v,k}|T_k>PL_v-b_v^{(k)} | H_k )P_{H_k}  \\
   & + \sum_{k=n+1,k\not\in \Omega_\text{const}}^{N_\text{fault modes}}P(|\mathbf{q}^{(k)}\bm{\varepsilon}|+T_{k,v}>PL_v -b_v^{(k)}| H_k) P_{H_k} \\
   & + \sum_{k\in \Omega_\text{const}}P(|(\hat{\mathbf{x}}^{(k)} - \mathbf{x})_{v}|+D_{k,v}>PL_v -b_v^{(k)}| H_k) P_{H_k} \\
& =  I_\text{REQ}^v\left(1- \frac{P_\text{not monitored}}{I_\text{REQ}} \right) \,.
\end{aligned}
\label{eq:I_cal_bound_add_const}
\end{equation}
\end{review}

Finally, with the equal allocation strategy on the integrity budget, the PL can be obtained by
\begin{review}
\begin{equation}
\begin{aligned}
    PL_v = \max \Biggl \{  & Q^{-1}_{(\hat{\mathbf{x}} - \mathbf{x})_v} \left(\frac{I_{REQ,0}^v}{2P_{H_0}}\right) + b_v^{(0)},  
    \max_{1<k\leq n} \left\{  Q^{-1}_{\mathbf{q}^{(k)}\bm{\varepsilon}} \left(\frac{I_{REQ,k}^v}{2P_{H_k}}\right) + |S_{v,k}|T_k + b_v^{(k)}\right\} ,\\
    & \max_{n<k\leq N_\text{fault modes},k\not\in \Omega_\text{const}} \left\{  Q^{-1}_{\mathbf{q}^{(k)}\bm{\varepsilon}} \left(\frac{I_{REQ,k}^v}{2P_{H_k}}\right) + T_{k,v} + b_v^{(k)}\right\} ,\\
& \max_{k\in \Omega_\text{const}} \left\{ \sigma_v^{(k)}Q^{-1}\left(\frac{I_{REQ,k}^v}{2P_{H_k}}\right) + D_{k,v} + b_v^{(k)}\right\} \Biggl\} \,.
\end{aligned}
\label{eq:PL_final_addconst}
\end{equation}
\end{review}

\section{Worldwide Simulation}\label{sec:simulation}
This section conducts a worldwide simulation to evaluate the performance of the proposed jackknife ARAIM algorithm.
The MATLAB Algorithm Availability Simulation Tool (MAAST) \citep{jan2001matlab} is utilized to simulate code ionosphere-free (IF) combination measurements with tropospheric correction, satellite positions, and user locations.
Both the single constellation (the nominal 24-satellite GPS constellation) and dual constellations (the aforementioned GPS constellations and the nominal 24-satellite Galileo constellation) cases are examined, where the almanacs file is defined in Table \ref{tb:almanacs}.
The users are placed on a grid every 15 degrees longitude and latitude (which gives 288 locations).
For each location, the geometries are simulated every 10 min (which gives 144 time steps).
The code IF combination measurements are simulated by adding a randomly generated sample from the given error distribution to the true range.
The proposed jackknife ARAIM algorithm is compared with the \textcolor{black}{SS ARAIM} algorithm \citep{joerger_solution_2014}.

\begin{table}[!htb]
\caption{Source of almanacs of the GPS and Galileo constellations}\label{tb:almanacs}
\centering
\begin{tblr}{
  width = \linewidth,
  colspec = {Q[120]Q[220]Q[370]},
hline{1,4} = {-}{1.5pt},
hline{2} = {-}{0.75pt},
}
Constellation & GPS Week of Almanacs & Source of Almanacs\\
GPS & 2243 & U.S.
Coast Guard Navigation Center \citep{GPS_almanac}\\
Galileo & 2243 & European GNSS Service Center \citep{Galileo_almanac}\\
\end{tblr}
\end{table}

The simulation of the measurement error distribution is detailed in Section \ref{subsec:nominal_error_simu_bound}.
Section \ref{subsec:single_worldwide} and Section \ref{subsec:dual_worldwide} give the results in the single-fault and multiple-fault scenarios, respectively.
\begin{review}Consistent with the proposed hybrid monitoring logic, single-fault modes are monitored using scalar jackknife residuals $t_k$, while each multi-fault mode is monitored using three axis-wise statistics $\tilde{t}_{k,v}$ ($v=1,2,3$) in the East/North/Up directions.\end{review}

\subsection{Nominal Error Simulation and Bounding}\label{subsec:nominal_error_simu_bound}
The measurement error of the code IF combination with respective to satellite $i$ and receiver $j$ consists the range projection of clock and orbit error $\varepsilon^i_{orb\&clk}$, tropospheric error $\varepsilon^i_{tropo,j}$, and multipath and code noise $\varepsilon^i_{\varrho,user,j,IF}$.
In \ref{app:SIS}, we use three-year ephemerides to characterize the normal performance of signal-in-space range error (SISRE) of GPS and Galileo satellites.
Results show that the SISRE of most satellites shows significant heavy-tailed properties.
Since the SISRE describes the statistical uncertainty of the modeled pseudorange due to errors in the broadcast orbit and clock information, the empirical distribution of SISRE is used to represent the distribution of the range projection of clock and orbit error $\varepsilon^i_{orb\&clk}$ in this experiment.
\par
The tropospheric error $\varepsilon^i_{tropo,j}$is assumed to have a zero-mean Gaussian distribution with the standard deviation given by RTCA-MOPS-229D \citep{RTCA20006}: 
\begin{equation}
\sigma^i_{tropo,j} = 0.12 [\text{m}] \frac{1.001}{\sqrt{0.002001+\sin^2{(\theta^i_j [\text{rad}])}}} \,,
    \label{eq:tropo_std}
\end{equation}
where $\theta^i_j$ is the elevation angle associated with the receiver $j$ and the satellite $i$.
The multipath and code noise $\varepsilon^i_{\varrho,user,j,IF}$ for airborne receivers is assumed to have a zero-mean Gaussian distribution with the same setting in \citep{blanch2015baseline}.
\par

The PDFs of the range projection of clock and orbit error, tropospheric error, and multipath and code noise are denoted as $f^i_{orb\&clk}(x)$, $f^i_{tropo,j}(x)$, and $f^i_{\varrho,user,j,IF}(x)$, respectively.
For each epoch, the nominal measurement error of the code IF combination is generated by summing up the randomly generated sample from $f^i_{orb\&clk}(x)$, $f^i_{tropo,j}(x)$, and $f^i_{\varrho,user,j,IF}(x)$, respectively.
Notably, $f^i_{orb\&clk}(x)$ is determined based on authentic experimental data instead of relying on empirical models.
This enhances the reliability of the experimental results obtained from the simulation.\par

Two types of nominal error bounds on the code IF combination can be obtained, including the non-Gaussian overbound $f^i_{\varrho,j,IF,acc}(x)$ and the Gaussian overbound $f^i_{\varrho,j,IF,Gaussian}(x)$ as follows:
\begin{subequations}
\begin{align}
    f^i_{\varrho,j,IF,acc}(x) &= f^i_{orb\&clk,PGO}(x) * f^i_{tropo,j,ob}(x) * f^i_{\varrho,user,j,IF,ob}(x) \\
    f^i_{\varrho,j,IF,Gaussian}(x) &= f^i_{orb\&clk,Gaussian}(x) * f^i_{tropo,j}(x) * f^i_{\varrho,user,j,IF}(x) \,,
\end{align}
\label{eq:bound_acc}
\end{subequations}
where $f^i_{orb\&clk,PGO}(x)$ and $f^i_{orb\&clk,Gaussian}(x)$ are the Principal Gaussian overbound (PGO) \citep{yan2024principal} and Gaussian overbound of the range projection of clock and orbit error, respectively.
The PGO is a non-Gaussian overbounding method, which provides a sharper yet conservative overbound than the Gaussian overbound for heavy-tailed error distributions \citep{yan2024principal}.
The parameters of the PGO and Gaussian overbound for each satellite are listed in Tables \ref{tb_GPS_bound} and \ref{tb_Galileo_bound} in \ref{app:SIS}.\par

The nominal error bounds in Eq.
\eqref{eq:bound_acc} are developed for accuracy evaluation and fault detection purposes.
For integrity purposes, the $b_{nom}$ term is introduced to create a symmetric error envelope based on the paired overbounding concept \citep{rife_paired_2004}.
The cumulative distribution function (CDF) of the nominal error bound for integrity can be written as follows:
\begin{subequations}
\begin{align}
    G^i_{\varrho,j,IF,int}(x) = \begin{cases}
        \int_{- \infty}^x f^i_{\varrho,j,IF,acc}(x+b_{nom,i}) \diff x & \text{if}~ G_v(x)<\frac{1}{2} \\
        \frac{1}{2} & \text{otherwise}\\ 
        \int_{- \infty}^x f^i_{\varrho,j,IF,acc}(x-b_{nom,i}) \diff x & \text{if}~ G_v(x)>\frac{1}{2} \\
    \end{cases} \\
    G^i_{\varrho,j,IF,Gaussianint}(x) = \begin{cases}
        \int_{- \infty}^x f^i_{\varrho,j,IF,Gaussian}(x+b_{nom,i}) \diff x & \text{if}~ G_v(x)<\frac{1}{2} \\
        \frac{1}{2} & \text{otherwise}\\ 
        \int_{- \infty}^x f^i_{\varrho,j,IF,Gaussian}(x-b_{nom,i}) \diff x & \text{if}~ G_v(x)>\frac{1}{2} \\
    \end{cases} \,,
\end{align}
\label{eq:bound_int}
\end{subequations}
where $G_v(x)$ is the empirical distribution of measurement errors of the code IF combination.
in \citet{walter2015keynote}, $b_{nom}$ is recommended to take $\SI{0.75}{\meter}$ to conservatively bound the bias impact.\par

In the experiment, the Gaussian overbound is used for the \textcolor{black}{SS ARAIM} algorithm.
For the jackknife ARAIM algorithm, both the Gaussian overbound and the non-Gaussian overbound are employed.
For notations, the jackknife ARAIM algorithm using the Gaussian overbound is named the JK-Gaussian ARAIM, while the one using the non-Gaussian overbound is named the JK-non-Gaussian ARAIM.
Table \ref{tb:RAIM_overbound_usage} lists the usage of overbounds in different ARAIM algorithms in the experiment.
\par
\begin{table}[!htb]
\caption{Overbounds used in different ARAIM algorithms}\label{tb:RAIM_overbound_usage}
\centering
\begin{tblr}{
  width = \linewidth,
  colspec = {Q[150]Q[200]Q[200]Q[200]},
hline{1,3} = {-}{1.5pt},
hline{2} = {-}{0.75pt},
}
Method & \textcolor{black}{SS ARAIM} & JK-Gaussian ARAIM & JK-non-Gaussian ARAIM \\
Overbounds & $f^i_{\varrho,j,IF,Gaussian}(x)$ & $f^i_{\varrho,j,IF,Gaussian}(x)$ & $f^i_{\varrho,j,IF,int}(x)$ &
\end{tblr}
\end{table}

\subsection{Single-Constellation Experiments}\label{subsec:single_worldwide}
In this section, the performance of the proposed JK-Gaussian ARAIM and JK-non-Gaussian ARAIM algorithms is evaluated in the single GPS constellation setting, where the \textcolor{black}{SS ARAIM} algorithm is taken as the benchmark.
The integrity and continuity budget and other parameters used for evaluating integrity monitoring algorithms are listed in Table \ref{tb:integrity_budgets}.
These values are aligned with the recommendation in the ARAIM algorithm description published by Worldwide GNSS Committee (WGC) \citep{RAIM_description_2017}.
The maximum number of simultaneous faults ($k_\text{max}$) that need to be monitored is determined by the method in \citet{blanch2015baseline}.
For the single constellation case, $k_\text{max}=1$.
For the dual constellation case, $k_\text{max}=2$.
An equal allocation strategy is adopted in allocating the integrity and continuity budgets to each fault mode.
\begin{table}[!htb]
\caption{Parameters used for evaluating integrity monitoring algorithms in the simulation}\label{tb:integrity_budgets}
\centering
\begin{tblr}{
  width = \linewidth,
  colspec = {Q[100]Q[400]Q[100]},
hline{1,9} = {-}{1.5pt},
hline{2} = {-}{0.75pt},
}
Parameter & Description  & Value \\
$I_\text{REQ}^3$ & Vertical integrity risk budget     & $9.8\times10^{-8}$     \\
$I_\text{REQ}^1+I_\text{REQ}^2$ & Horizontal integrity risk budget    & $2\times10^{-9}$   \\
$C_\text{REQ,FA}^3$ & Vertical continuity risk budget allocated to false alarms   &  $3.9\times10^{-6}$    \\
$C_\text{REQ}^1+C_\text{REQ}^2$ & Horizontal continuity risk budget allocated to false alarms &  $9\times10^{-8}$     \\
$P_\text{sat}$ & Prior probability of satellite fault per approach  &  $10^{-5}$     \\
$P_\text{const}$ & Prior probability of constellation fault per approach  &  $10^{-4}$     \\
$P_\text{THRES}$ & Threshold for the integrity risk coming from unmonitored faults & $9\times10^{-8}$
\end{tblr}
\end{table}

\begin{figure}[!htb]
    \centering
  \subfloat[~~~~]{%
\includegraphics[width=75mm]{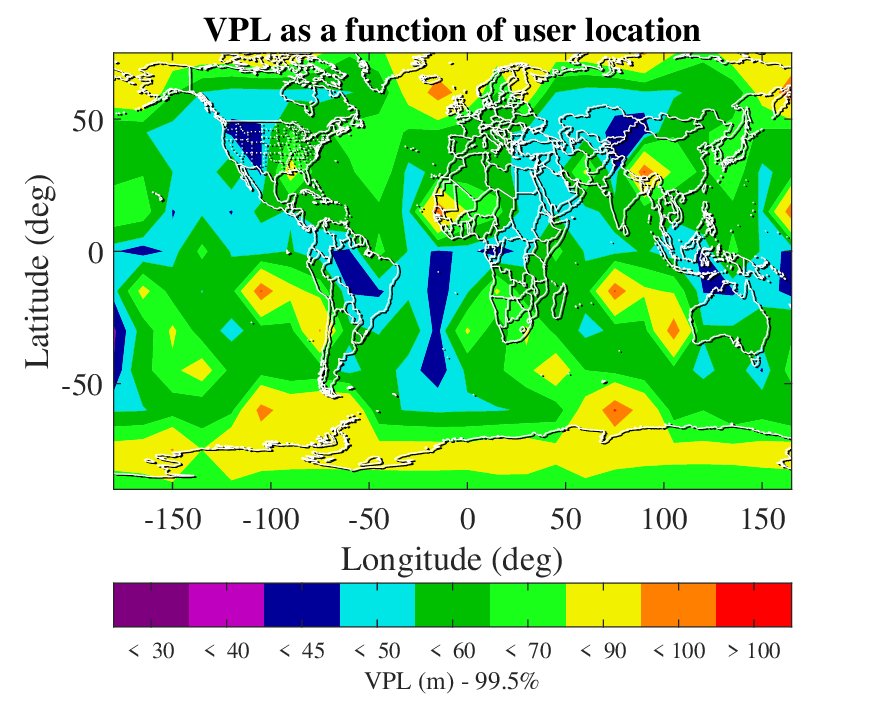}}
    \hfill
  \subfloat[~~~~]{%
\includegraphics[width=75mm]{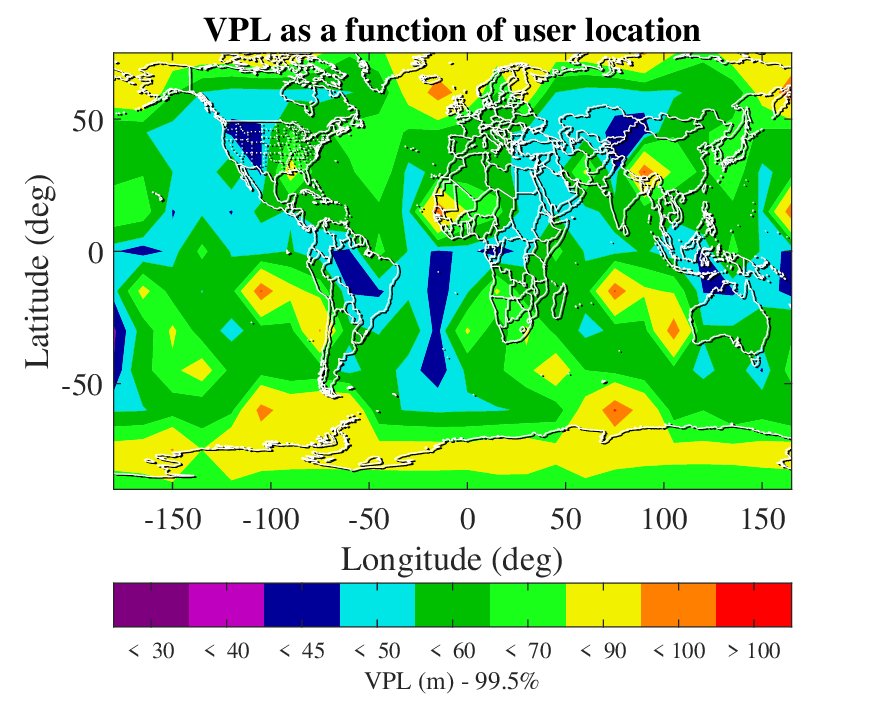}}
    \\
  \subfloat[~~~~]{%
\includegraphics[width=75mm]{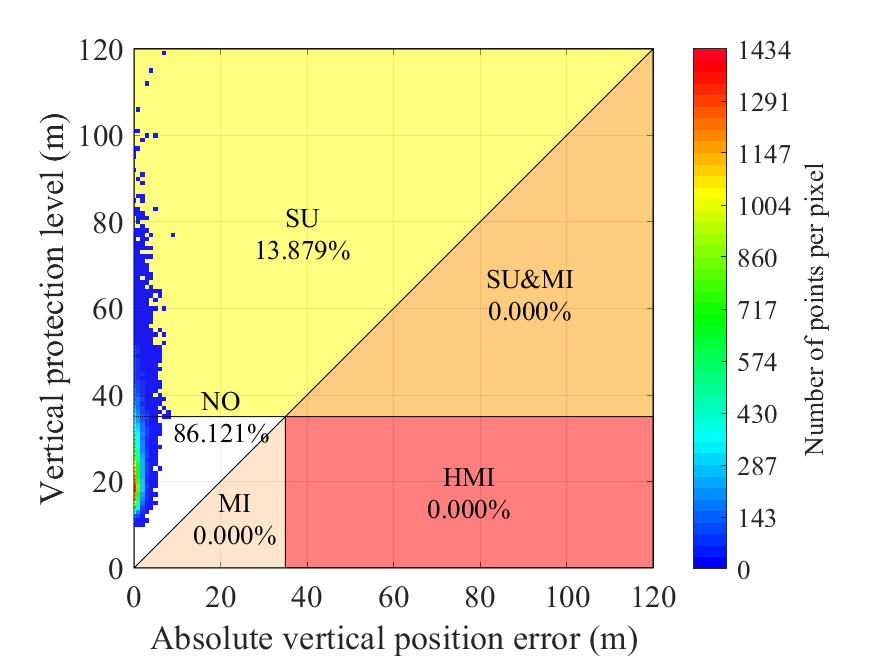}}
    \hfill
  \subfloat[~~~~]{%
\includegraphics[width=75mm]{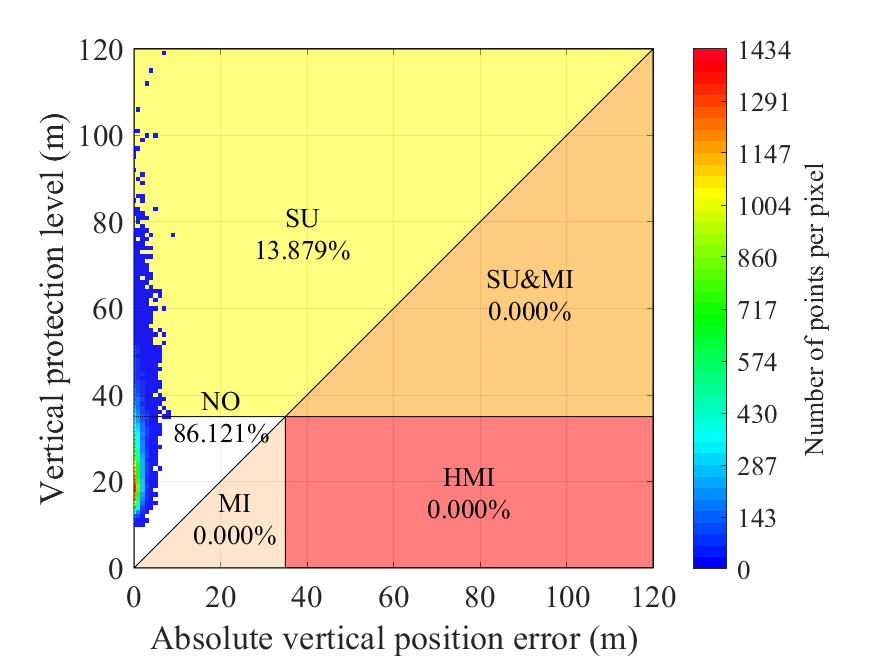}}
\caption{99.5 percentile of the VPL over the course of the day yielded by (a) the \textcolor{black}{SS ARAIM} and (b) the proposed JK-Gaussian ARAIM for the single constellation; and the triangular chart of (c) the \textcolor{black}{SS ARAIM} and (d) the proposed JK-Gaussian ARAIM regarding the vertical performance for the single constellation. ``NO" represents normal operation, ``MI" represents misleading information, ``SU" represents system unavailable, ``SU\&MI" represents system unavailable and misleading information, and ``HMI" represents hazardously misleading information.}
  \label{fig:SF_PL_RealSimu_Gaussian_singleConst}
\end{figure}

The first analysis involves the comparison between the \textcolor{black}{SS ARAIM} and the proposed JK-Gaussian ARAIM algorithms, both of which use the Gaussian overbound for code IF combination nominal errors.
Fig.
\ref{fig:SF_PL_RealSimu_Gaussian_singleConst}a and Fig.
\ref{fig:SF_PL_RealSimu_Gaussian_singleConst}b show the map of 99.5 percentile of the VPL over the course of a day of the \textcolor{black}{SS ARAIM} and the proposed JK-Gaussian ARAIM algorithms, respectively.
As can be seen, the two methods yield the same results, where the 99.5 percentile VPL is larger than \SI{50}{\meter} in most user locations.\par

To gain a comprehensive understanding of the performance of the two methods, the triangular charts of the \textcolor{black}{SS ARAIM} and the JK-Gaussian ARAIM regarding the vertical performance are plotted in Figure \ref{fig:SF_PL_RealSimu_Gaussian_singleConst}c and Figure \ref{fig:SF_PL_RealSimu_Gaussian_singleConst}d, respectively, which again demonstrates the equivalence of the two methods.
Specifically, each bin in the triangular chart represents the number of occurrences of a specific pair of absolute vertical positioning error (VPE) and VPL among all $288\times144$ location-time events.
The percentage of the normal operation (the VPL is larger than the VPE but less than the vertical alert limit (VAL), i.e., \SI{35}{\meter} here) is around \SI{86}{\percent}.
The percentage of misleading information (the VPE is larger than the VPL but less than the VAL) and hazardously misleading information (the VPE is larger than the VAL without alerts) events are all zero for both methods.\par

\begin{figure}[!htb]
    \centering
  \subfloat[~~~~]{%
\includegraphics[width=75mm]{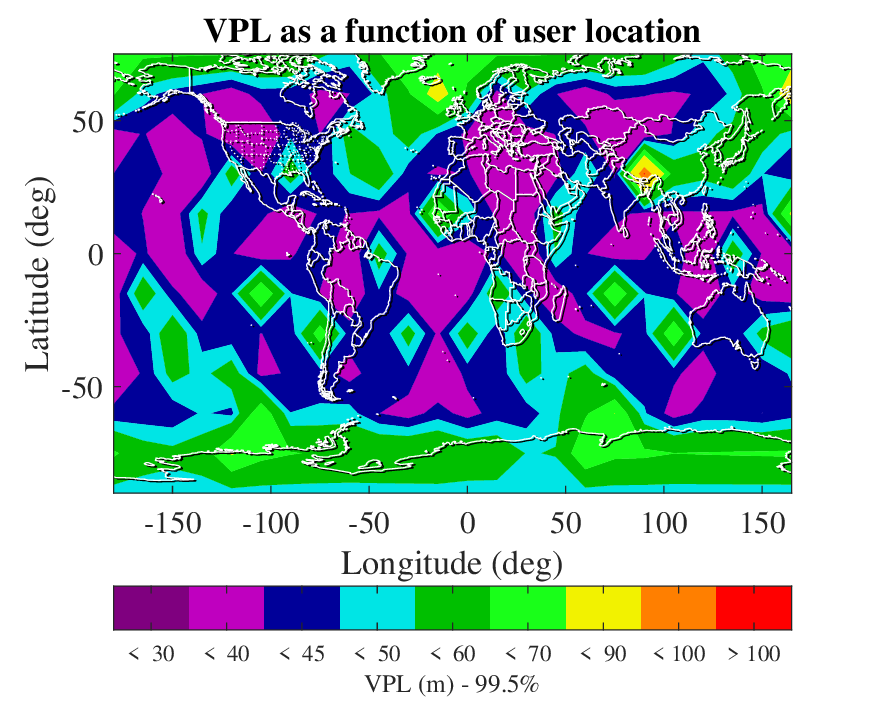}}
    \hfill
  \subfloat[~~~~]{%
\includegraphics[width=75mm]{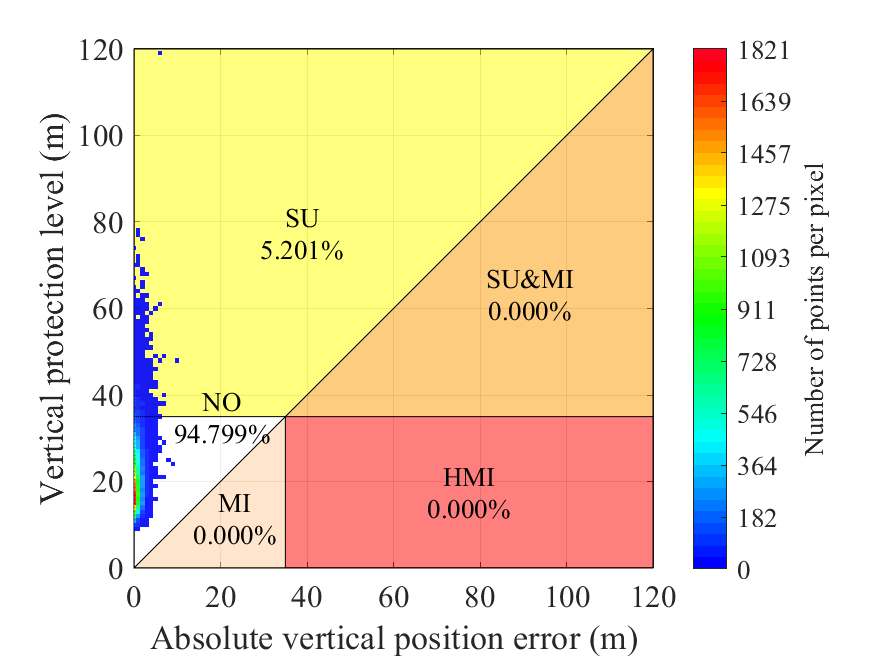}}
\caption{(a) 99.5 percentile of the VPL over the course of the day yielded by the proposed JK-non-Gaussian ARAIM for the single constellation; (b) The triangular chart of the proposed JK-non-Gaussian ARAIM regarding the vertical performance for the single constellation.}
  \label{fig:SF_PL_RealSimu_nonGaussian_singleConst}
\end{figure}

The second analysis focuses on the additional benefits brought by introducing non-Gaussian overbound into the jackknife ARAIM algorithm.
Fig.
\ref{fig:SF_PL_RealSimu_nonGaussian_singleConst}a shows the map of 99.5 percentile of the VPL over the course of a day of the proposed JK-non-Gaussian ARAIM algorithm.
As can be seen, the 99.5 percentile VPL is less than \SI{45}{\meter} in most user locations.
By comparing with the results in Fig.
\ref{fig:SF_PL_RealSimu_Gaussian_singleConst}b, one can conclude that introducing a non-Gaussian overbound into the jackknife ARAIM algorithm can further reduce the VPL.
The triangular chart of the JK-non-Gaussian ARAIM in Fig.
\ref{fig:SF_PL_RealSimu_nonGaussian_singleConst}b further confirms this conclusion, where the distribution of the VPE-VPL pairs shows a higher concentration level than that of the jackknife ARAIM algorithm and the \textcolor{black}{SS ARAIM} algorithm.
More importantly, the percentage of the normal operation of the JK-non-Gaussian ARAIM method increases to \SI{94.799}{\percent}, indicating that the JK-non-Gaussian ARAIM seldom compromises integrity.
\par

For a better understanding of the possibility of using the JK-non-Gaussian ARAIM to support LPV-200 precision approach operations, Table \ref{tb:coverage_singleConst} summarizes the coverage of the three methods with $VAL=\SI{35}{\meter}$ at different levels of system availability.
The system availability is the fraction of time that VPL is less than a given VAL at a given location, while the coverage is the fraction of the earth that satisfies a given system availability.
All three methods show satisfactory performance in coverage under \SI{75}{\percent} system availability.
However, when the availability requirements increase to \SI{95}{\percent},  the \textcolor{black}{SS ARAIM} and the JK-Gaussian ARAIM algorithms only have a coverage of \SI{15.16}{\percent}.
In contrast, the coverage of the JK-non-Gaussian ARAIM still keeps above \SI{88}{\percent} in this condition.
Nevertheless, the coverage of the JK-non-Gaussian ARAIM decreases to \SI{7.84}{\percent} under \SI{99.5}{\percent} system availability.
The above results reveal that the proposed JK-non-Gaussian ARAIM method has huge potential to support integrity applications with harsh navigation requirements.

\begin{table}[!htb]
\caption{Coverage for the single constellation at different levels of system availability}\label{tb:coverage_singleConst}
\centering
\begin{tblr}{
  width = \linewidth,
  colspec = {Q[100]Q[150]Q[200]Q[200]Q[250]},
  cell{2}{1} = {r=2}{},
hline{1,5} = {-}{1.5pt},
hline{2} = {-}{0.75pt},
vline{3} = {-}{0.75pt},
}
VAL & Availability & \textcolor{black}{SS ARAIM} & JK-Gaussian ARAIM & JK-non-Gaussian ARAIM \\
\textcolor{black}{\SI{35}{\meter}} & \SI{75}{\percent}   & \SI{96.3}{\percent}    & \SI{96.3}{\percent}           & \textbf{100}~\si{\percent}     \\
& \SI{95}{\percent} & \SI{15.16}{\percent}  & \SI{15.16}{\percent}   & \textbf{88.64}~\si{\percent} \\
& \SI{99.5}{\percent}   & \SI{0}{\percent}    & \SI{0}{\percent}          & \textbf{7.84}~\si{\percent}
\end{tblr}
\end{table}

\subsection{Dual-Constellation Experiments}\label{subsec:dual_worldwide}
\begin{figure}[!htb]
    \centering
  \subfloat[~~~~]{%
\includegraphics[width=75mm]{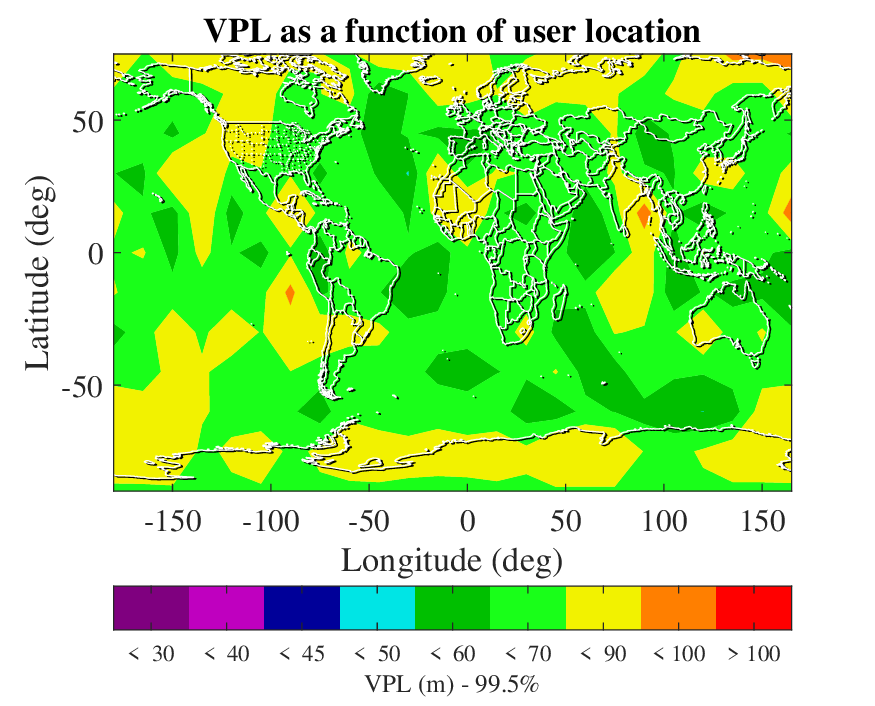}}
    \hfill
  \subfloat[~~~~]{%
\includegraphics[width=75mm]{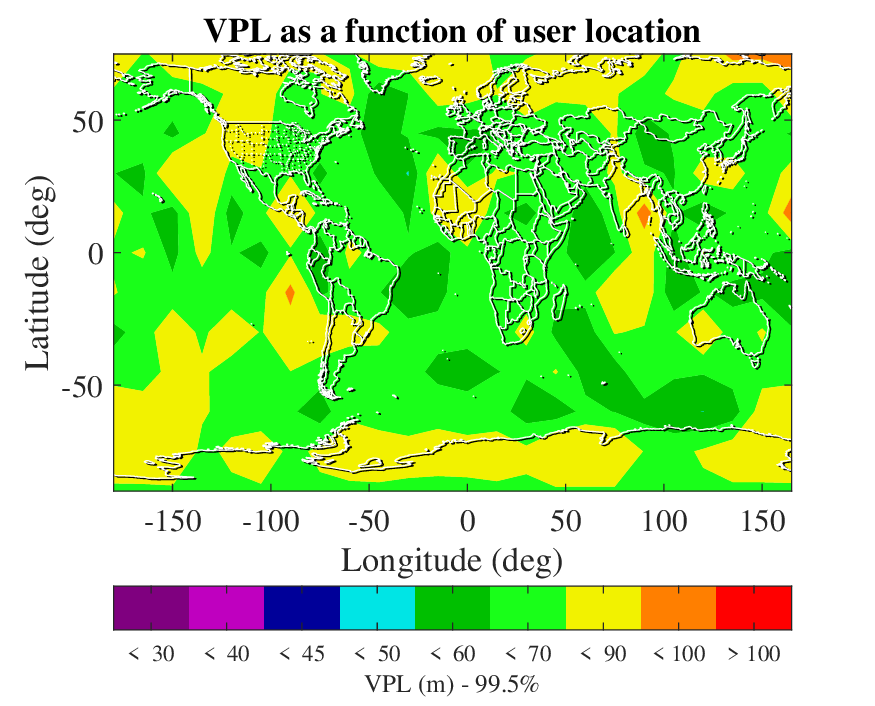}}
    \\
  \subfloat[~~~~]{%
\includegraphics[width=75mm]{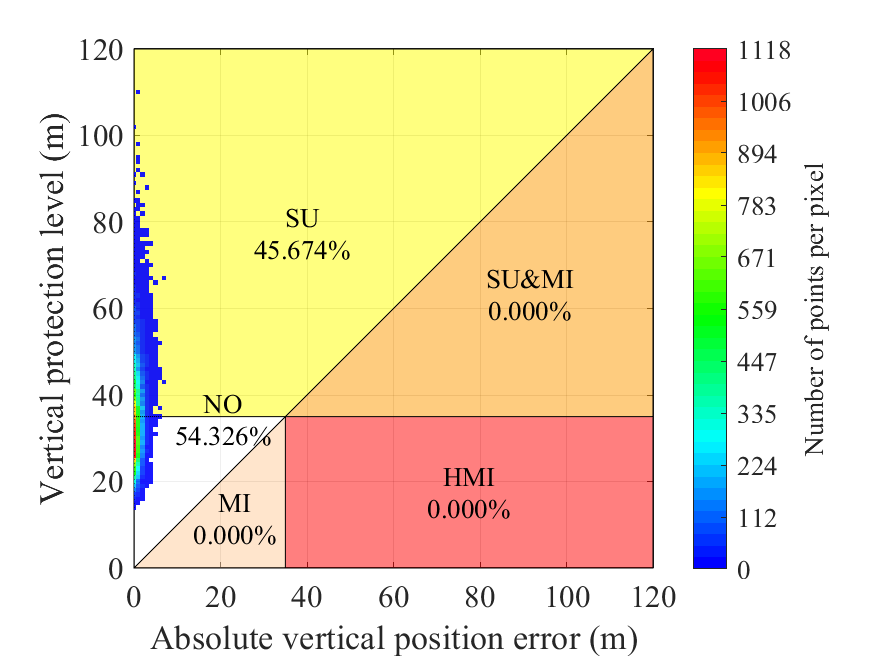}}
    \hfill
  \subfloat[~~~~]{%
\includegraphics[width=75mm]{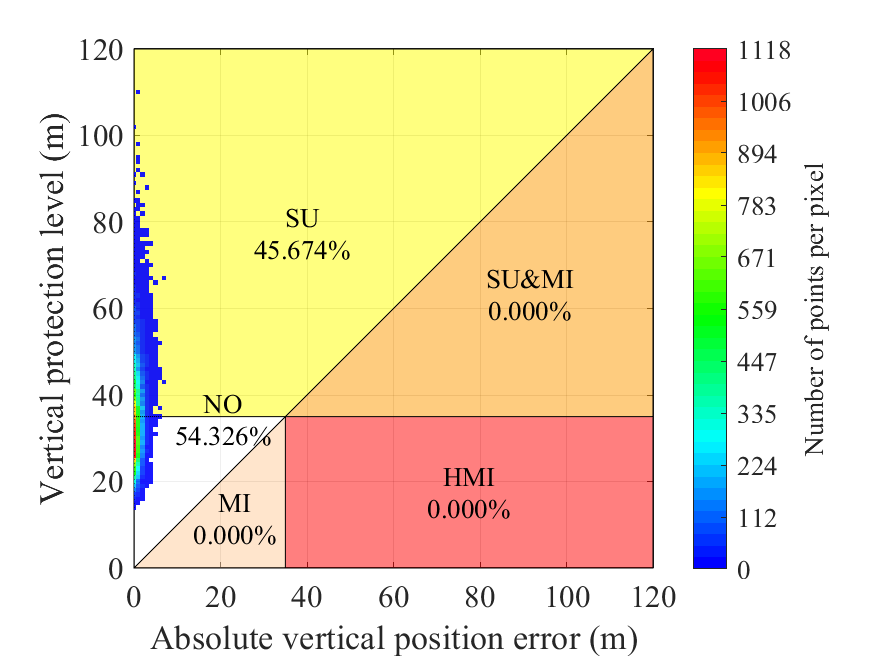}}
\caption{99.5 percentile of the VPL over the course of the day yielded by (a) the \textcolor{black}{SS ARAIM} and (b) the proposed JK-Gaussian ARAIM for the dual constellation; and the triangular chart of (c) the \textcolor{black}{SS ARAIM} and (d) the proposed JK-Gaussian ARAIM regarding the vertical performance for the dual constellation. }
  \label{fig:MF_PL_RealSimu_Gaussian_dualConst}
\end{figure}

This section evaluates the performance of the proposed JK-Gaussian ARAIM and JK-non-Gaussian ARAIM algorithms in the dual constellation setting.
The simulation parameters are given in Table \ref{tb:integrity_budgets}.
Similar to the single constellation setting in Section \ref{subsec:single_worldwide}, the JK-Gaussian ARAIM exhibits the equivalent performance to the \textcolor{black}{SS ARAIM}, as shown in the 99.5 percentile VPL map in Fig.
\ref{fig:MF_PL_RealSimu_Gaussian_dualConst}a and Fig.
\ref{fig:MF_PL_RealSimu_Gaussian_dualConst}b.
However, the magnitude of the 99.5 percentile VPL of these two methods exceeds \SI{60}{\meter} at most user locations, which is significantly larger than that in the single constellation setting (see Fig.
\ref{fig:SF_PL_RealSimu_nonGaussian_singleConst}a and Fig.
\ref{fig:SF_PL_RealSimu_nonGaussian_singleConst}b).
This is because the SISRE of Galileo satellites in the dual constellation setting has significant heavy-tailed properties (as revealed in \ref{app:SIS}), which results in the over-conservatism in the finalized Gaussian overbounds of code IF combination errors.
Such conservatism is passed to the position domain bounding, eventually enlarging the VPLs in the dual-constellation setting.
As a consequence, the system unavailability events of both methods experience a surge in the dual-constellation setting, which can be observed in the triangular chart in Figures \ref{fig:MF_PL_RealSimu_Gaussian_dualConst}c and \ref{fig:MF_PL_RealSimu_Gaussian_dualConst}d, where the system unavailability events with $VAL=\SI{35}{\meter}$ account for \SI{45.674}{\percent}.
\par

\begin{figure}[!htb]
    \centering
  \subfloat[~~~~]{%
\includegraphics[width=75mm]{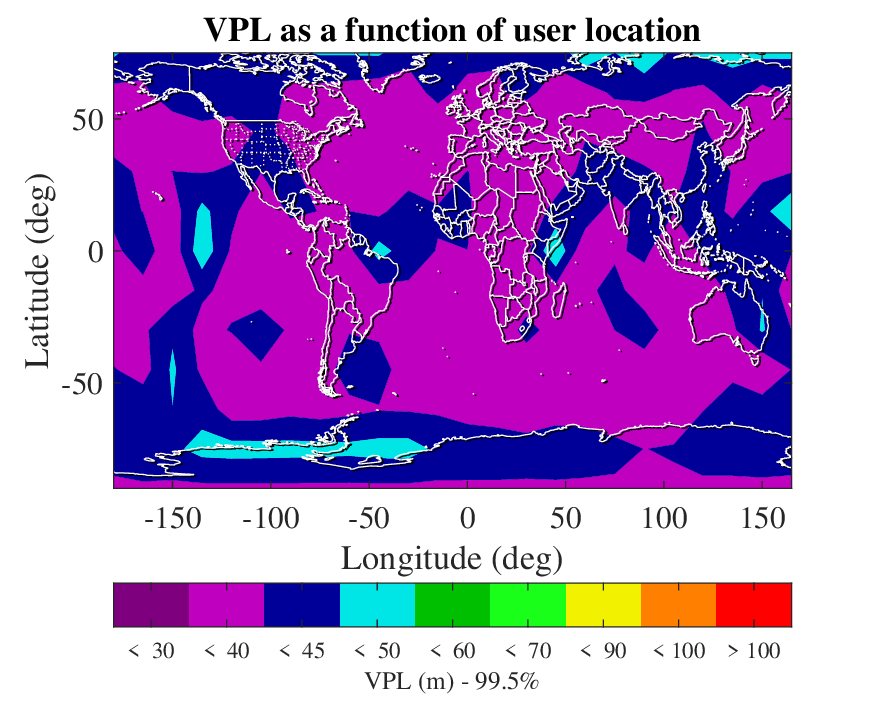}}
    \hfill
  \subfloat[~~~~]{%
\includegraphics[width=75mm]{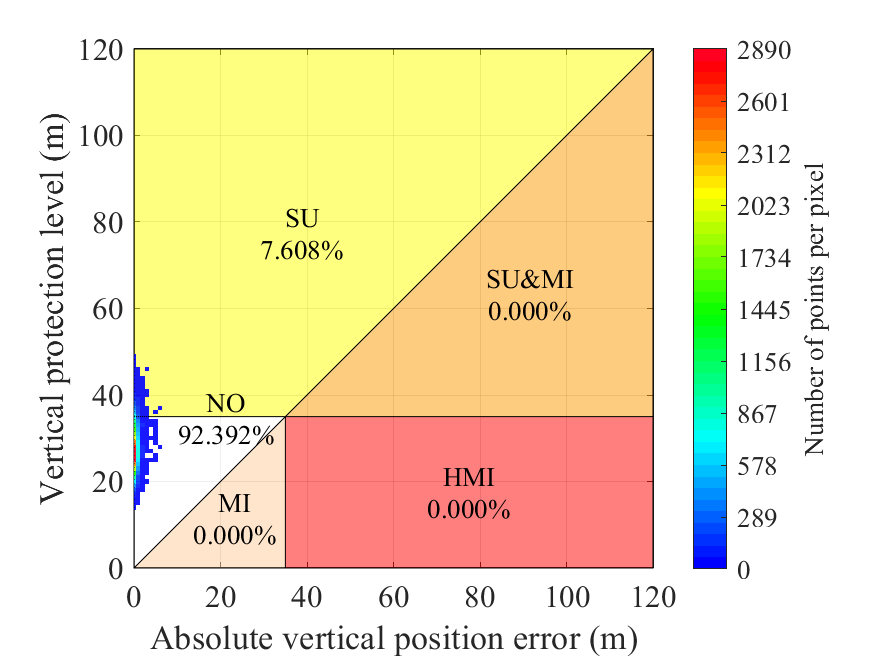}}
\caption{(a) 99.5 percentile of the VPL over the course of the day yielded by the proposed JK-non-Gaussian ARAIM for the dual constellation; (b) The triangular chart of the proposed JK-non-Gaussian ARAIM regarding the vertical performance for the dual constellation.}
  \label{fig:MF_PL_RealSimu_nonGaussian_dualConst}
\end{figure}

Nevertheless, the JK-non-Gaussian ARAIM still shows satisfactory performance in the dual-constellation setting, where the 99.5 percentile VPL is smaller than \SI{40}{\meter} in most user locations (Fig.
\ref{fig:MF_PL_RealSimu_nonGaussian_dualConst}a) and the VPE-VPL pairs have extremely concentrated distribution (Fig.
\ref{fig:MF_PL_RealSimu_nonGaussian_dualConst}b).
Moreover, the percentage of the normal operation events with $VAL=\SI{35}{\meter}$ even exceeds \SI{92}{\percent}, making it possible to support LPV-200 precision approach operations \citep{icao2006Annex10}.
\par

Table \ref{tb:coverage_dualConst} summarizes the coverage of the three methods with $VAL=\SI{35}{\meter}$ at different levels of system availability.
The \textcolor{black}{SS ARAIM} and the JK-Gaussian ARAIM have a \SI{54}{\percent} coverage even under \SI{75}{\percent} system availability.
This result is expected because both the \textcolor{black}{SS ARAIM} and JK-Gaussian ARAIM use an over-conservative Gaussian overbound.
In contrast, the coverage of the JK-non-Gaussian ARAIM is nearly \SI{100}{\percent} under \SI{75}{\percent} system availability.
Its coverage even exceed \SI{62}{\percent} under \SI{95}{\percent} system availability.
These results reveal the huge potential of the JK-non-Gaussian ARAIM algorithm to support LPV-200 requirements using the GPS-Galileo dual constellation.
\par

\begin{table}[!htb]
\caption{Coverage for the dual constellation at different levels of system availability}\label{tb:coverage_dualConst}
\centering
\begin{tblr}{
  width = \linewidth,
  colspec = {Q[100]Q[150]Q[200]Q[200]Q[250]},
  cell{2}{1} = {r=2}{},
hline{1,5} = {-}{1.5pt},
hline{2} = {-}{0.75pt},
vline{3} = {-}{0.75pt},
}
VAL & Availability & \textcolor{black}{SS ARAIM} & JK-Gaussian ARAIM & JK-non-Gaussian ARAIM \\
\textcolor{black}{\SI{35}{\meter}} &\SI{75}{\percent}    & \SI{54}{\percent} & \SI{54}{\percent} & \textbf{99.29}~\si{\percent}   \\
& \SI{95}{\percent}   & \SI{0}{\percent}     & \SI{0}{\percent}    & \textbf{62.55}~\si{\percent}  \\
& \SI{99.5}{\percent} & \SI{0}{\percent}     & \SI{0}{\percent} & \textbf{3.68}~\si{\percent}
\end{tblr}
\end{table}

It is worth noting that the reporting result about the \textcolor{black}{SS ARAIM} in this simulation study is quite different from the findings in \citet{blanch2010raim,joerger_fault_2016}, from which the \textcolor{black}{SS ARAIM} is examined to be able to provide global coverage for LPV-200 in the GPS-Galileo dual constellation.
The primary reason is that these studies use hypothetical models to simulate the range errors, which results in over-optimistic results.
For example, the 1-sigma error bound of Galileo SISRE is set to be \SI{0.96}{\meter} in \citet{joerger_fault_2016}, which is significantly smaller than the value determined by experimental data in Table \ref{tb_Galileo_bound} in \ref{app:SIS}.
In such a condition, the system availability of \textcolor{black}{SS ARAIM} is overestimated.
\par

\section{Computation Time Comparison}\label{sec:computation}
To evaluate the computational efficiency of the proposed jackknife ARAIM method, we conducted two experiments with different constellation configurations.
All computations were performed on a laptop equipped with an Intel Core i7-12700H CPU running at 2.30GHz.

\begin{figure}[!htb]
  \centering
\subfloat[~~~~]{%
\includegraphics[width=60mm]{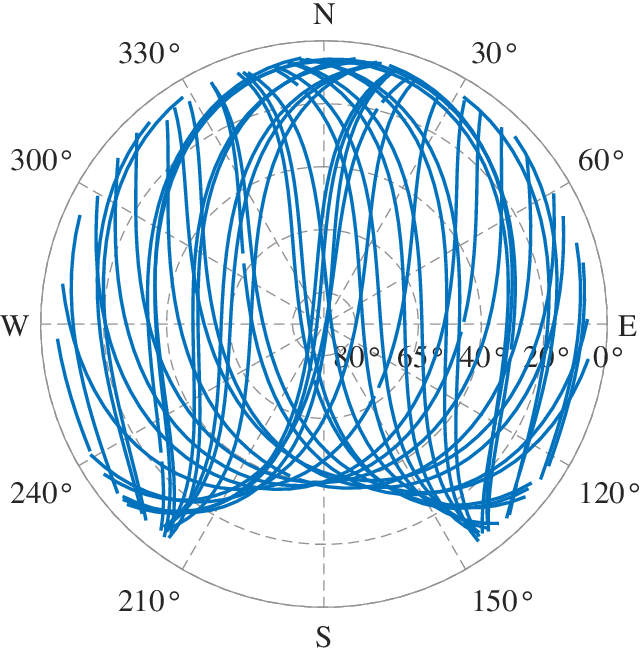}}
  \hfill
\subfloat[~~~~]{%
\includegraphics[width=75mm]{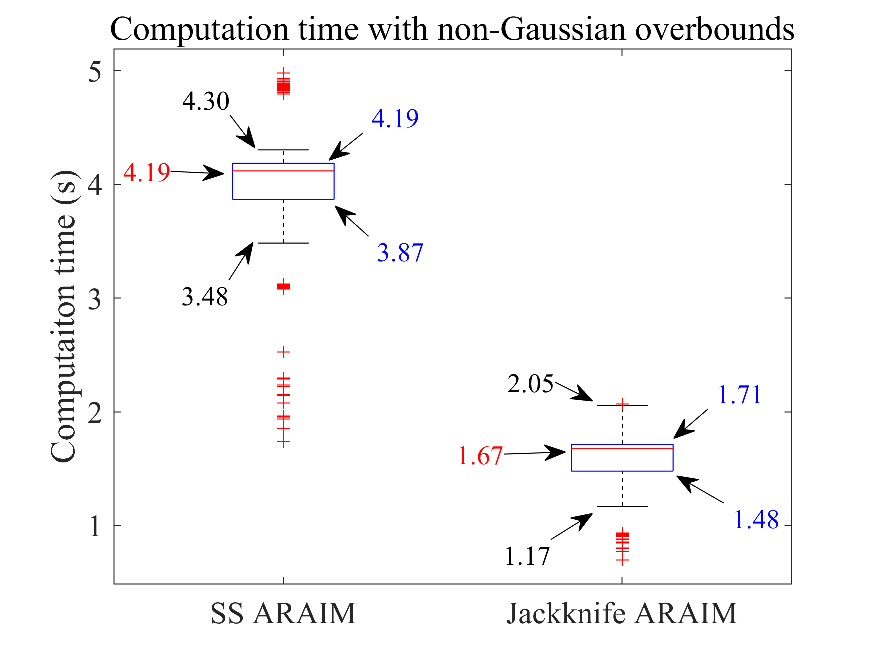}}
\caption{Computation time comparison for GPS-only constellation: (a) Trajectories of GPS satellites over a 24-hour period; (b) The box plot of computation times for jackknife and \textcolor{black}{SS ARAIM} methods both with non-Gaussian overbounds.}
\label{fig:computation_GPS}
\end{figure}

In the first experiment, we considered a GPS-only constellation scenario with the same configuration as in Section \ref{subsec:single_worldwide}.
A user location was randomly selected at $(15^\circ S,120^\circ E,0 m)$, and satellite geometries were simulated at 10-minute intervals over a 24-hour period, resulting in 144 time steps.
The code IF combination measurements and their corresponding nominal errors were simulated following the methodology described in Section \ref{subsec:nominal_error_simu_bound}.
Both the jackknife ARAIM and \textcolor{black}{SS ARAIM} methods were implemented with non-Gaussian overbounds at each time step, using the integrity monitoring parameters specified in Table \ref{tb:integrity_budgets}.
Figure \ref{fig:computation_GPS}a illustrates the trajectory of GPS satellites over the 24-hour period, while Figure \ref{fig:computation_GPS}b presents a box plot comparing the computation times of both methods.
The results demonstrate a substantial reduction in computational complexity achieved by the jackknife ARAIM method.
Specifically, the jackknife ARAIM exhibits a median computation time of \SI{1.67}{second}, compared to \SI{4.11}{second} for the \textcolor{black}{SS ARAIM} method, representing a \SI{59.4}{\percent} reduction in processing time.
This significant improvement in computational efficiency suggests the potential applicability of the jackknife ARAIM method in real-time safety-critical GNSS integrity monitoring applications.

\begin{figure}[!htb]
  \centering
\subfloat[~~~~]{%
\includegraphics[width=60mm]{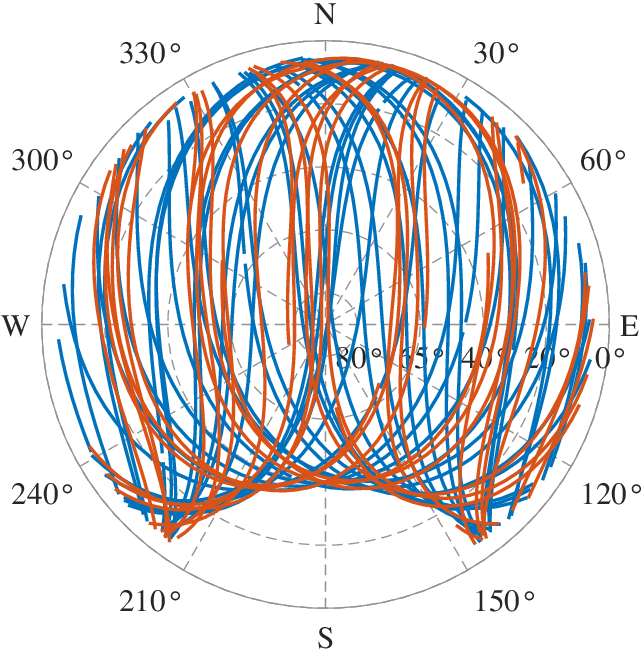}}
  \hfill
\subfloat[~~~~]{%
\includegraphics[width=75mm]{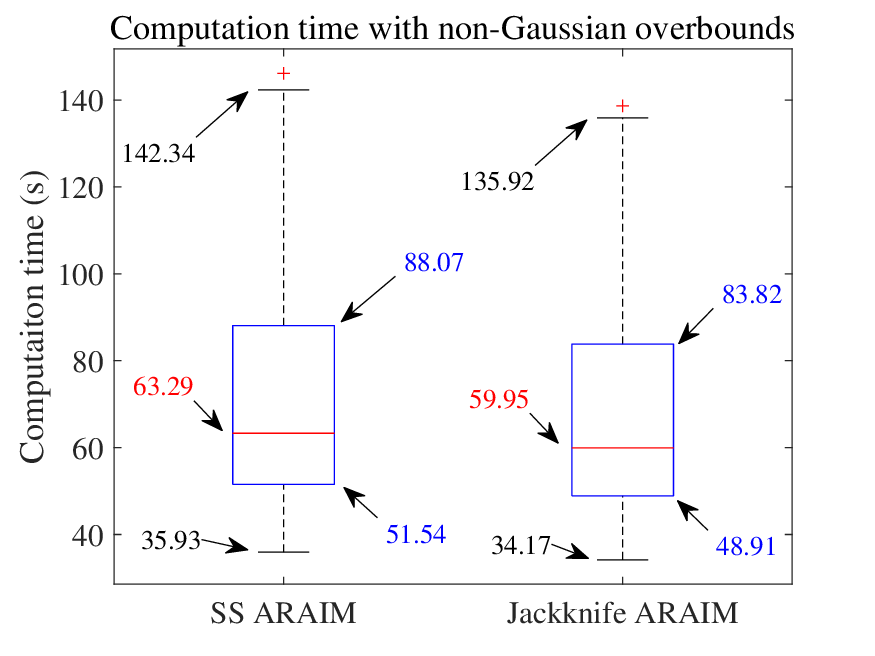}}
\caption{Computation time comparison for GPS-Galileo dual-constellation: (a) Trajectories of GPS satellites over a 24-hour period; (b) The box plot of computation times for jackknife and \textcolor{black}{SS ARAIM} methods both with non-Gaussian overbounds.}
\label{fig:computation_GPS_Galileo}
\end{figure}

\begin{review}We conducted a second experiment using a combined GPS-Galileo constellation.
All other experimental parameters remained identical to the first experiment.
Figure \ref{fig:computation_GPS_Galileo}a shows the trajectories of both GPS and Galileo satellites over the 24-hour period, and Figure \ref{fig:computation_GPS_Galileo}b presents the computation time comparison. As can be seen, the jackknife ARAIM exhibits a median computation time of \SI{59.95}{second}, compared to \SI{63.30}{second} for the \textcolor{black}{SS ARAIM} method, representing only a marginal \SI{5.3}{\percent} reduction in processing time. This result indicates that the computational efficiency advantage of the jackknife ARAIM method is primarily realized in single-fault cases, as demonstrated in the GPS-only experiment. In multiple-fault scenarios, which are more prevalent in dual-constellation configurations, both algorithms exhibit similar computational efficiency since they both employ axis-wise (East/North/Up) monitoring for multi-fault modes.\end{review}

\begin{review}The remarkable computational advantage of the proposed jackknife ARAIM method in single-fault scenarios can be fundamentally attributed to the scalar nature of the jackknife-based monitoring statistics for single-fault modes.\end{review}
Our previous theoretical analysis has demonstrated that the jackknife detector, which utilizes these scalar test statistics, can achieve approximately \begin{review}threefold~\end{review} enhancement in computational efficiency compared to the solution separation detector.
\begin{review}In this work, the proposed jackknife ARAIM method inherits this efficiency through a hybrid monitoring strategy: it preserves scalar tests for single-fault modes while using only three axis-wise (East/North/Up) monitors for each multi-fault mode, thereby matching MHSS monitoring dimensionality without constructing high-dimensional solution-separation statistics for every hypothesis. This explains why the computational efficiency advantage is primarily realized in single-fault cases, while in multiple-fault scenarios, both algorithms exhibit similar efficiency since they both employ axis-wise monitoring for multi-fault modes.\end{review}

\section{Conclusions and Future work}\label{sec_conclusions}

This paper extends the jackknife detector to simultaneous fault detection with non-Gaussian nominal errors.
It is proven that the constructed test statistic is the linear combination of measurement errors without making assumptions about the distribution of errors, which provides an accurate probabilistic model for hypothesis testing.
An integrity monitoring algorithm for multi-constellation GNSS navigation is further developed by systematically exploiting the properties of the jackknife detector in the range domain.
A tight bound of the integrity risk is derived by quantifying the impacts of hypothetical fault vectors on the position solution.
The performance of the proposed integrity monitoring algorithm is evaluated through a worldwide simulation with both single GPS and GPS-Galileo dual constellation settings.
Results show that the proposed method has the equivalent performance with the \textcolor{black}{SS ARAIM} algorithm using the same nominal error models.
\begin{review}However, the proposed method demonstrates remarkable computational efficiency in single-fault scenarios, achieving a \SI{59.4}{\percent} reduction in processing time for single-constellation scenarios compared to \textcolor{black}{SS ARAIM}. In dual-constellation scenarios, where multiple-fault cases are more prevalent, both algorithms exhibit similar computational efficiency since they both employ axis-wise (East/North/Up) monitoring for multi-fault modes\end{review}.
Despite the GPS and Galileo constellations, the proposed method is also applicable to other constellations, such as BeiDou and GLObalnaya NAvigatsionnaya Sputnikovaya Sistema in Russian (GLONASS).
By incorporating these additional constellations, the system availability can be further improved.
However, additional efforts are needed to characterize the nominal error performance of satellites in these constellations, which is out of the scope of this study.
\par

This study has several limitations, which also point out future research directions.
Similar to the \textcolor{black}{SS ARAIM} method, the Bonferroni correction is applied to the jackknife ARAIM to handle multi-testing problems.
However, the Bonferroni correction is overly conservative, which can raise miss-detection risks.
A possible remedy is to apply the Holm–Bonferroni correction \citep{holm1979simple}, which keeps the family-wise error rate no higher than a pre-specified significance level.
However, Holm–Bonferroni correction involves the systematic adjustment of the significance level for each individual test.
It is essential to investigate and remove the impacts of such adjustments on system integrity.
In addition, the proposed algorithm mainly focuses on fault detection and is designed to raise alarms when faults are detected.
Future work can improve the jackknife ARAIM by incorporating fault exclusion processes.
Additional tests must be devised to monitor wrong exclusions and include these effects in protection level calculations, ensuring that any performance gains do not compromise overall system safety.

\section*{Declarations}
\subsection*{Conflict of Interest}
The authors declare that they have no competing interests.

\subsection*{Data Availability Statement}
Not applicable.

\subsection*{Funding Statement}
Not applicable.

\subsection*{Author Contribution}
Methodology, P.Y, J.Z; Experiment design, P.Y, R.J; Writing the original draft, P.Y, L.H, R.J; Review and editing, R.J, J.Z, C.W; All authors read and approved the manuscript.

\appendix
\renewcommand{\thesection}{Appendix \Alph{section}}
\newpage

\section{Relationships with Multiple Hypothesis Solution Separation}\label{sec:app_SS_JK}
\textbf{Equivalence in single-fault case.} Let 
\begin{equation}
    d_{k,q} = (\hat{\mathbf{x}}- \hat{\mathbf{x}}^{(k)})_q, q=1,2,3,\cdots,m \,,
\label{eq:SS}
\end{equation}
be the separation between the full solution and the $k$th subsolution  \citep{brenner1996integrated, blanch2010raim,blanch2015baseline}, where the subscript $q$ represents the $q$th component of the solution.
Define the separation vector $\mathbf{d}_k$ as follows:
\begin{equation}
\mathbf{d}_k = [d_{k,1},d_{k,2},d_{k,3},\cdots,d_{k,m}]^T .
\end{equation}
\citet{yan2025efficient} identifies the following one-to-one maping relationship between jackknife test statistics $t_k$ and the solution seperation vector $\mathbf{d}_k$
\begin{equation}
\mathbf{d}_k = (\mathbf{G}^T \mathbf{W}\mathbf{G})^{-1} \mathbf{g}_k^T W_{k,k} t_k \,.
    \label{eq:ss_jk_relation}
\end{equation}

This relationship reveals that the difference in solutions, induced by excluding the $k$th measurement, is proportional to the perturbation of the $k$th measurement and oriented along the derivative of the solution $\hat{\mathbf{x}}$ with respect to the $k$-th measurement.
Moreover, Eq.
\eqref{eq:ss_jk_relation} indicates mapping from a scalar to a high-dimensional vector, revealing the difference in computation load between the solution separation and jackknife detectors during hypothesis testing.
In a single-constellation system, the computational load of the SS detector is \begin{review}three~\end{review} times that of the jackknife detector \citep{yan2025efficient}.\par

\begin{review}
\noindent\textbf{Extension to simultaneous faults (multiple-fault modes).}
We now show that the above equivalence extends to the fault mode $k$ with an excluded index set $idx_k^{ex}$ (with $|idx_k^{ex}|\geq 1$), which is the case addressed in Section~\ref{subsec:reconstruct_JK}.
Without loss of generality (by permuting the measurement ordering), partition the measurement model in Eq.~\eqref{eq:general_linear} as
\begin{equation}
    \mathbf{y} =
    \begin{bmatrix}
    \mathbf{y}_{in}\\
    \mathbf{y}_{ex}
    \end{bmatrix},~~
    \mathbf{G} =
    \begin{bmatrix}
    \mathbf{G}_{in}\\
    \mathbf{G}_{ex}
    \end{bmatrix},~~
    \mathbf{W}=
    \begin{bmatrix}
    \mathbf{W}_{in} & \mathbf{0}\\
    \mathbf{0} & \mathbf{W}_{ex}
    \end{bmatrix},
    \label{eq:partition_in_ex}
\end{equation}
where the block ``$ex$'' collects measurements with indices in $idx_k^{ex}$ and the block ``$in$'' collects all other measurements.
The full-set information matrix is $\mathbf{N}=\mathbf{G}^T\mathbf{W}\mathbf{G}$, while the subset (fault-mode-$k$) information matrix is $\mathbf{N}_{in}=\mathbf{G}_{in}^T\mathbf{W}_{in}\mathbf{G}_{in}$.
Since $\mathbf{W}$ is diagonal, we have the decomposition
\begin{equation}
    \mathbf{N} = \mathbf{N}_{in} + \mathbf{G}_{ex}^T\mathbf{W}_{ex}\mathbf{G}_{ex}\,.
    \label{eq:N_decomp}
\end{equation}
By applying the Woodbury matrix identity to Eq.~\eqref{eq:N_decomp}, one obtains
\begin{equation}
    \mathbf{N}^{-1}
    = \mathbf{N}_{in}^{-1} - \mathbf{N}_{in}^{-1}\mathbf{G}_{ex}^T
    \left(\mathbf{W}_{ex}^{-1}+\mathbf{G}_{ex}\mathbf{N}_{in}^{-1}\mathbf{G}_{ex}^T\right)^{-1}
    \mathbf{G}_{ex}\mathbf{N}_{in}^{-1}\,.
    \label{eq:woodbury_multifault}
\end{equation}
The full-set solution matrix $\mathbf{S}$ is given by
\begin{equation}
    \mathbf{S} = (\mathbf{G}^T\mathbf{W}\mathbf{G})^{-1}\mathbf{G}^T\mathbf{W} = \mathbf{N}^{-1}\mathbf{G}^T\mathbf{W}\,.
    \label{eq:S_derivation}
\end{equation}
The subset solution of fault mode $k$ is given by
\begin{equation}
    \hat{\mathbf{x}}^{(k)} = \left(\mathbf{G}_{in}^T\mathbf{W}_{in}\mathbf{G}_{in}\right)^{-1}\mathbf{G}_{in}^T\mathbf{W}_{in}\mathbf{y}_{in}
    =\mathbf{N}_{in}^{-1}\mathbf{G}_{in}^T\mathbf{W}_{in}\mathbf{y}_{in}\,.
    \label{eq:subsolution_in}
\end{equation}
From the partition in Eq.~\eqref{eq:partition_in_ex}, the matrix $\mathbf{G}^T\mathbf{W}$ can be written as
\begin{equation}
    \mathbf{G}^T\mathbf{W}=\left[\mathbf{G}_{in}^T\mathbf{W}_{in}~~\mathbf{G}_{ex}^T\mathbf{W}_{ex}\right]\,,
    \label{eq:GtW_partition}
\end{equation}
where the first block corresponds to included measurements and the second block to excluded measurements.
Substituting Eq.~\eqref{eq:GtW_partition} into Eq.~\eqref{eq:S_derivation}, we have
\begin{equation}
    \mathbf{S} = \mathbf{N}^{-1}\left[\mathbf{G}_{in}^T\mathbf{W}_{in}~~\mathbf{G}_{ex}^T\mathbf{W}_{ex}\right]\,.
\end{equation}
Let $\mathbf{S}_{ex}$ denote the columns of $\mathbf{S}$ corresponding to the excluded set $idx_k^{ex}$, i.e.,
\begin{equation}
    \mathbf{S}_{ex} = \mathbf{N}^{-1}\mathbf{G}_{ex}^T\mathbf{W}_{ex}\,.
    \label{eq:S_ex_def}
\end{equation}
Define
\begin{equation}
    \mathbf{M} \triangleq \mathbf{W}_{ex}^{-1}+\mathbf{G}_{ex}\mathbf{N}_{in}^{-1}\mathbf{G}_{ex}^T\,.
    \label{eq:M_def_multifault}
\end{equation}
By substituting Eq.~\eqref{eq:woodbury_multifault} into Eq.~\eqref{eq:S_ex_def}, we have
\begin{equation}
\begin{aligned}
    \mathbf{S}_{ex}
    &=\left(\mathbf{N}_{in}^{-1}-\mathbf{N}_{in}^{-1}\mathbf{G}_{ex}^T\mathbf{M}^{-1}\mathbf{G}_{ex}\mathbf{N}_{in}^{-1}\right)\mathbf{G}_{ex}^T\mathbf{W}_{ex}\\
    &=\mathbf{N}_{in}^{-1}\mathbf{G}_{ex}^T\left[\mathbf{I}-\mathbf{M}^{-1}\left(\mathbf{G}_{ex}\mathbf{N}_{in}^{-1}\mathbf{G}_{ex}^T\right)\right]\mathbf{W}_{ex}\,.
\end{aligned}
    \label{eq:S_ex_expand_multifault}
\end{equation}
Using Eq.~\eqref{eq:M_def_multifault}, we have $\mathbf{G}_{ex}\mathbf{N}_{in}^{-1}\mathbf{G}_{ex}^T=\mathbf{M}-\mathbf{W}_{ex}^{-1}$, and thus
\begin{equation}
    \mathbf{I}-\mathbf{M}^{-1}\left(\mathbf{G}_{ex}\mathbf{N}_{in}^{-1}\mathbf{G}_{ex}^T\right)
    =\mathbf{I}-\mathbf{M}^{-1}\left(\mathbf{M}-\mathbf{W}_{ex}^{-1}\right)
    =\mathbf{M}^{-1}\mathbf{W}_{ex}^{-1}\,.
    \label{eq:bracket_identity_multifault}
\end{equation}
Substituting Eq.~\eqref{eq:bracket_identity_multifault} into Eq.~\eqref{eq:S_ex_expand_multifault} and using $\mathbf{W}_{ex}^{-1}\mathbf{W}_{ex}=\mathbf{I}$ yields the compact identity
\begin{equation}
    \mathbf{S}_{ex}
    = \mathbf{N}_{in}^{-1}\mathbf{G}_{ex}^T \mathbf{M}^{-1}
    = \mathbf{N}_{in}^{-1}\mathbf{G}_{ex}^T\left(\mathbf{W}_{ex}^{-1}+\mathbf{G}_{ex}\mathbf{N}_{in}^{-1}\mathbf{G}_{ex}^T\right)^{-1}.
    \label{eq:S_ex_identity}
\end{equation}
Now expand the full-set solution $\hat{\mathbf{x}}=\mathbf{S}\mathbf{y}$:
\begin{equation}
    \hat{\mathbf{x}} = \mathbf{N}^{-1}\left(\mathbf{G}_{in}^T\mathbf{W}_{in}\mathbf{y}_{in}+\mathbf{G}_{ex}^T\mathbf{W}_{ex}\mathbf{y}_{ex}\right)\,.
    \label{eq:full_sol_start}
\end{equation}
Substituting the Woodbury expansion of $\mathbf{N}^{-1}$ from Eq.~\eqref{eq:woodbury_multifault} into the first term, and using $\mathbf{S}_{ex}=\mathbf{N}^{-1}\mathbf{G}_{ex}^T\mathbf{W}_{ex}$ from Eq.~\eqref{eq:S_ex_def} for the second term, we have
\begin{equation}
\begin{aligned}
    \hat{\mathbf{x}}
    &= \left(\mathbf{N}_{in}^{-1}-\mathbf{N}_{in}^{-1}\mathbf{G}_{ex}^T\mathbf{M}^{-1}\mathbf{G}_{ex}\mathbf{N}_{in}^{-1}\right)\mathbf{G}_{in}^T\mathbf{W}_{in}\mathbf{y}_{in}+\mathbf{S}_{ex}\mathbf{y}_{ex} \\
    &= \mathbf{N}_{in}^{-1}\mathbf{G}_{in}^T\mathbf{W}_{in}\mathbf{y}_{in}-\mathbf{N}_{in}^{-1}\mathbf{G}_{ex}^T\mathbf{M}^{-1}\mathbf{G}_{ex}\mathbf{N}_{in}^{-1}\mathbf{G}_{in}^T\mathbf{W}_{in}\mathbf{y}_{in}+\mathbf{S}_{ex}\mathbf{y}_{ex}\,.
\end{aligned}
    \label{eq:full_sol_intermediate}
\end{equation}
Substitute Eq.~\eqref{eq:S_ex_identity} and Eq.~\eqref{eq:subsolution_in} into Eq.~\eqref{eq:full_sol_intermediate}, we have
\begin{equation}
\begin{aligned}
    \hat{\mathbf{x}}
    &= \hat{\mathbf{x}}^{(k)}-\mathbf{S}_{ex}\mathbf{G}_{ex}\hat{\mathbf{x}}^{(k)}+\mathbf{S}_{ex}\mathbf{y}_{ex} \\
    &= \hat{\mathbf{x}}^{(k)} + \mathbf{S}_{ex}\left(\mathbf{y}_{ex}-\mathbf{G}_{ex}\hat{\mathbf{x}}^{(k)}\right)\,.
\end{aligned}
    \label{eq:full_sol_expand_multifault}
\end{equation}
The jackknife residual vector of the excluded measurements under fault mode $k$ is
\begin{equation}
    \mathbf{t}_{ex}^{(k)} = \mathbf{y}_{ex}-\mathbf{G}_{ex}\hat{\mathbf{x}}^{(k)}\,,
    \label{eq:t_ex_def}
\end{equation}
whose elements are the jackknife residuals $t_i^{(k)},i\in idx_k^{ex}$,defined in Eq.~\eqref{eq:define_statistic}.
Combining Eqs.~\eqref{eq:full_sol_expand_multifault}--\eqref{eq:t_ex_def} yields the vector identity
\begin{equation}
    \hat{\mathbf{x}}-\hat{\mathbf{x}}^{(k)} = \mathbf{S}_{ex}\mathbf{t}_{ex}^{(k)}\,.
    \label{eq:ss_jk_relation_multifault_vec}
\end{equation}
Taking the $v$th component of Eq.~\eqref{eq:ss_jk_relation_multifault_vec} gives
\begin{equation}
    (\hat{\mathbf{x}}-\hat{\mathbf{x}}^{(k)})_v = \sum_{i\in idx_k^{ex}} S_{v,i}t_i^{(k)}\,,
    \label{eq:ss_jk_relation_multifault}
\end{equation}
which shows that the SS statistic $(\hat{\mathbf{x}}-\hat{\mathbf{x}}^{(k)})_v$ is identical to the multi-fault jackknife statistic $\tilde{t}_{k,v}$ in Eq.~\eqref{eq:combint_JK}.
\end{review}

\section{Signal-In-Space Range Error and Bounding}\label{app:SIS}
SISRE describes the statistical uncertainty of the modeled pseudorange due to errors in the broadcast orbit and clock information \citep{montenbruck2015broadcast, perea2017ura, walter2010evaluation}.
Satellite orbit and clock errors arise due to uncertainties in the Orbit Determination and Time Synchronization (ODTS) process managed by the Constellation Service Providers (CSP) \citep{perea2017ura}.
A common method to evaluate broadcast orbit and clock errors is to calculate the deviations between the satellite's position and clock bias, which is provided by the broadcast ephemeris (BCE) and the precise ephemeris (PCE) \citep{montenbruck2015broadcast,montenbruck2018multi}.
In this work, the BCE is acquired from the International GNSS Service (IGS) BRDC files in RINEX format (Version 3) for both GPS and Galileo.
The PCE is obtained from the Center for Orbit Determination in Europe (CODE), with sampling intervals of 15 minutes for GPS satellites and 5 minutes for Galileo satellites.
Following the same method in \citet{walter2018validation}, which defines the SISRE as the user projected error (UPE), we evaluate the nominal performance of GPS SISRE with respect to L1/L2 combination over a three-year period from January 1st, 2020, to December 31st, 2022.
The analysis for Galileo satellites is conducted with respect to the E1/E5a combination within the same period.

\subsection{Nominal Performance Characterization}\label{app:SISRE_performance}
\begin{figure}[!htb]
    \centering
  \subfloat[~~~~]{%
\includegraphics[width=75mm]{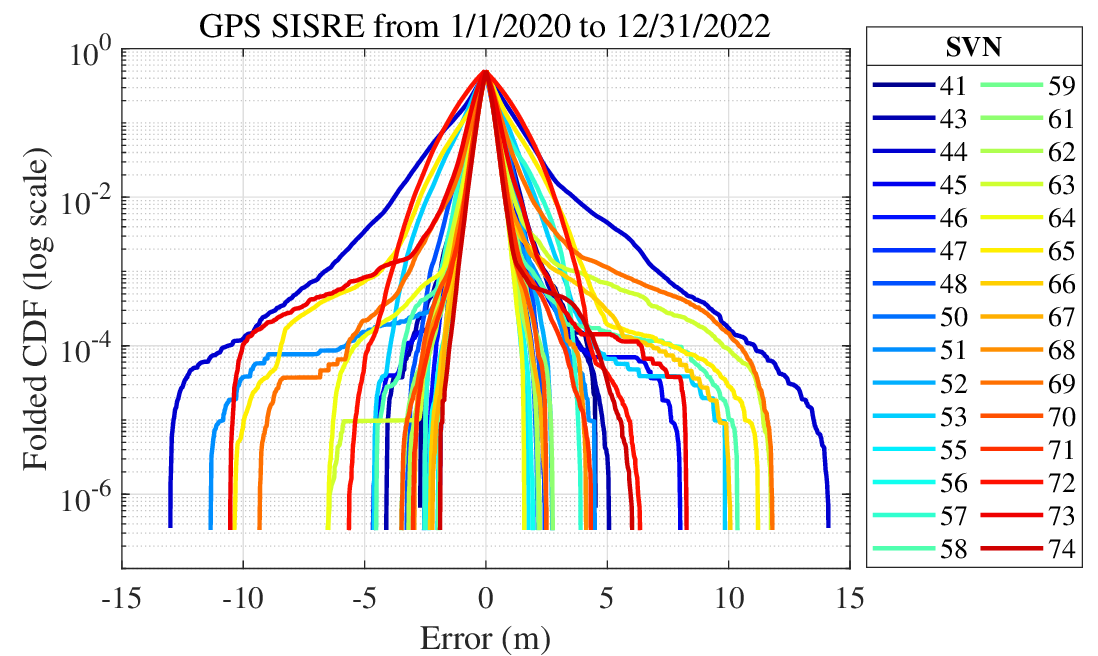}}
    \hfill
  \subfloat[~~~~]{%
\includegraphics[width=75mm]{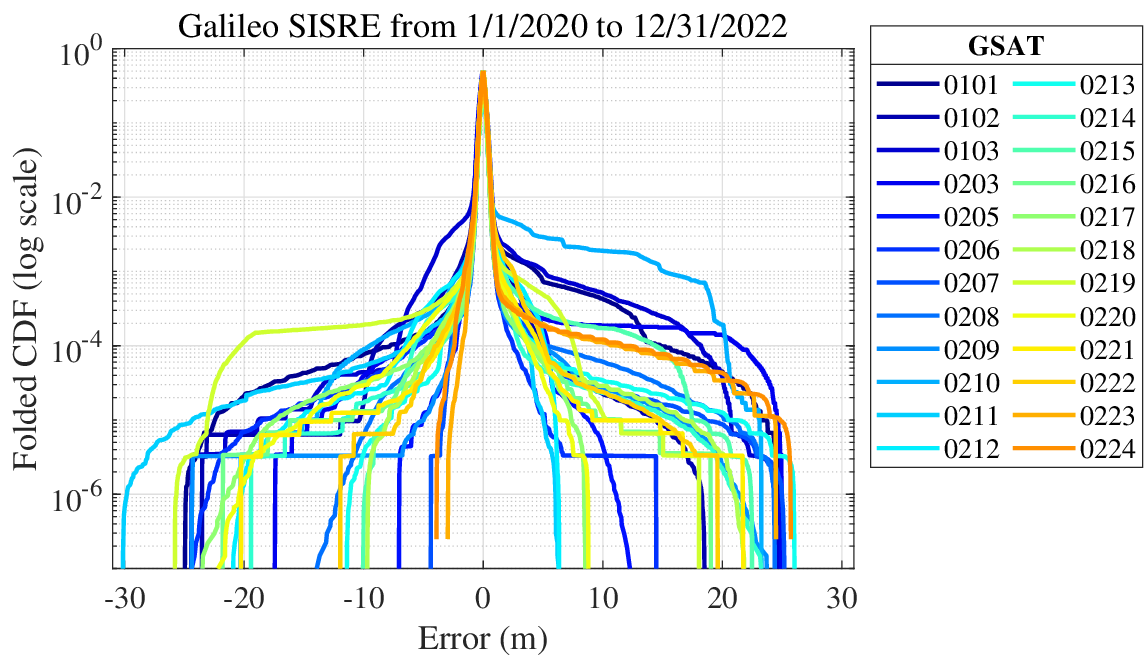}}
\caption{The folded CDF of (a) GPS and (b) Galileo $SISRE_{UPE}$ for individual satellites from January 1st, 2020 to December 31st, 2022.}
  \label{fig:UPE_all_individual_3year}
\end{figure}

Fig.
\ref{fig:UPE_all_individual_3year}a plots the folded CDF of $SISRE_{UPE}$ for each GPS satellite, where significant differences among satellites are observed.
Some satellites, such as SVN 44, SVN 51, SVN 73, and SVN 65, exhibit large error magnitude and dispersion, with their maximum $SISRE_{UPE}$ exceeding \SI{10}{\meter}.
However, the $SISRE_{UPE}$ of most satellites is relatively small, which retains within the range of $\pm \SI{5}{\meter}$.
Table \ref{tb_GPS_bound} summarizes the standard deviation of the $SISRE_{UPE}$ for each satellite, which also suggests the difference among satellites.
The mean of $SISRE_{UPE}$ for each satellite is also listed in Table \ref{tb_GPS_bound}, with the magnitude less than \SI{5}{\cm} for most satellites.
Three categories of $SISRE_{UPE}$ distributions can be identified as follows: 1) Two-side heavy-tailed $SISRE_{UPE}$; 2) One-side heavy-tailed $SISRE_{UPE}$; and 3) Gaussian-liked $SISRE_{UPE}$.
The category information is also provided in Table \ref{tb_GPS_bound}.\par

The folded CDF of $SISRE_{UPE}$ for individual Galileo satellites is depicted in Fig.
\ref{fig:UPE_all_individual_3year}b.
Two categories of $SISRE_{UPE}$ distributions can be identified as follows: 1) Two-side heavy-tailed $SISRE_{UPE}$ and 2) One-side heavy-tailed $SISRE_{UPE}$.
Intuitively speaking, the tailedness of the Galileo $SISRE_{UPE}$ is much heavier than that of the GPS $SISRE_{UPE}$.
However, the statistics of Galileo $SISRE_{UPE}$ in Table \ref{tb_Galileo_bound} suggest that the standard deviation of the Galileo $SISRE_{UPE}$ is relatively smaller than that of the GPS $SISRE_{UPE}$.
These findings suggest that Galileo satellites usually have smaller SISRE than GPS satellites, but Galileo satellites have larger worse-case nominal SISRE.
Finally, another important piece of information in Table \ref{tb_Galileo_bound} is that the mean value of the Galileo $SISRE_{UPE}$ is nearly zero, which is similar to the GPS case.

\subsection{Bounding Signal-In-Space Range Error}
Two overbounding methods, including the Gaussian overbound \citep{decleene_defining_2000} and the Principal Gaussian overbound \citep{yan2024principal}, are employed to bound GPS and Galileo SISRE.
The latter one is a non-Gaussian overbounding method.

\subsubsection{Gaussian overbound} \label{app:tsgo}
Let the CDF of the random variable $v$ be $G_v$.
The Gaussian overbound is determined by finding the minimum $\delta$ that satisfies
\begin{subequations}
\begin{align}
    {\int_{- \infty}^{x}{f_\mathcal{N}\left( {x;0,\delta} \right)}}dx {\geq}& G_v(x)~\forall x < 0 \\
      {\int_{- \infty}^{x}{f_\mathcal{N}\left( {x;0,\delta} \right)}}dx {\leq}& G_v(x)~\forall x \geq 0 \,,
    \label{eq_TSGO_right}
\end{align}
\end{subequations}
where $f_\mathcal{N}(x;0,\sigma)$ is the PDF of a zero-mean Gaussian distribution with a standard deviation of $\sigma$.
\subsubsection{Principal Gaussian overbound}\label{app:pgo}
The Principal Gaussian overbound \citep{yan2024principal} utilizes the zero-mean bimodal Gaussian mixture model (BGMM) to fit the error distribution based on the expectation–maximization (EM) algorithm \citep{dempster_maximum_1977} and divides the BGMM into the core and tail regions based on the analysis of BGMM membership weight.
Within each region, one of the Gaussian components in the BGMM holds a dominant position, and a CDF overbound is constructed based on the dominant Gaussian component.
The PDF of the Principal Gaussian overbound (PGO) is given by
\begin{equation}
    f_{PGO}(x) = \begin{cases}
        \left( {1 + k} \right)\left( {1 - p_{1}} \right)f_N\left( {x;0,\sigma_{2}} \right) &|x| > x_{rp} \\
        p_{1}f_N\left( {x;0,\sigma_{1}} \right) + c &|x| \leq x_{rp} 
    \end{cases} \,,
\label{eq_PGO}
\end{equation}
where $f_N\left( {x;0,\sigma_{1}} \right)$ and $f_N\left( {x;0,\sigma_{2}} \right)$ are the PDF of the first and the second Gaussian component of the fitted BGMM, $\sigma_{1}$ and $\sigma_{2}$ are the corresponding standard deviations, and $p_1$ and $1-p_1$ are the mixing weight of the two Gaussian components, respectively; $k$, $c$, and $x_{rp}$ are parameters uniquely determined by the partition strategy based on the analysis of BGMM membership weight \citep{yan2024principal}.
\par

A detailed description of PGO can refer to \citep{yan2024principal}.
Soon, it will be shown in \ref{app:bounding_performance} that PGO provides a sharper yet conservative overbound than the Gaussian overbound for heavy-tailed error distribution.
Notably, it is proven that PGO can maintain the overbounding property through convolution \citep{yan2024principal}, which is the basis for deriving pseudorange-level requirements from the position domain integrity requirements \citep{decleene_defining_2000}.
\par

\subsection{Bounding Performance of SISRE}\label{app:bounding_performance}
\begin{figure}[!htb]
    \centering
  \subfloat[]{%
\includegraphics[width=55mm]{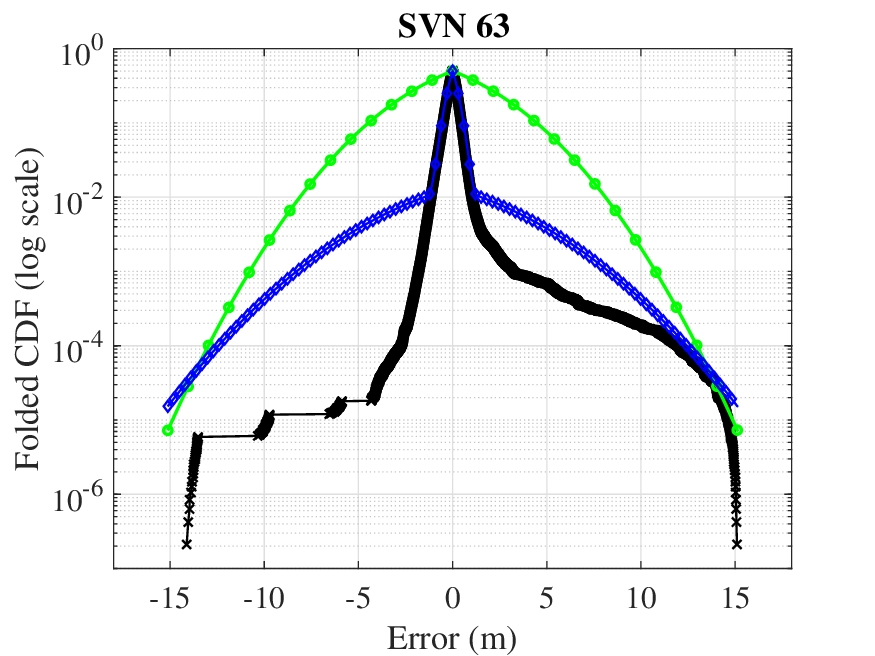}}
   \subfloat[]{%
\includegraphics[width=55mm]{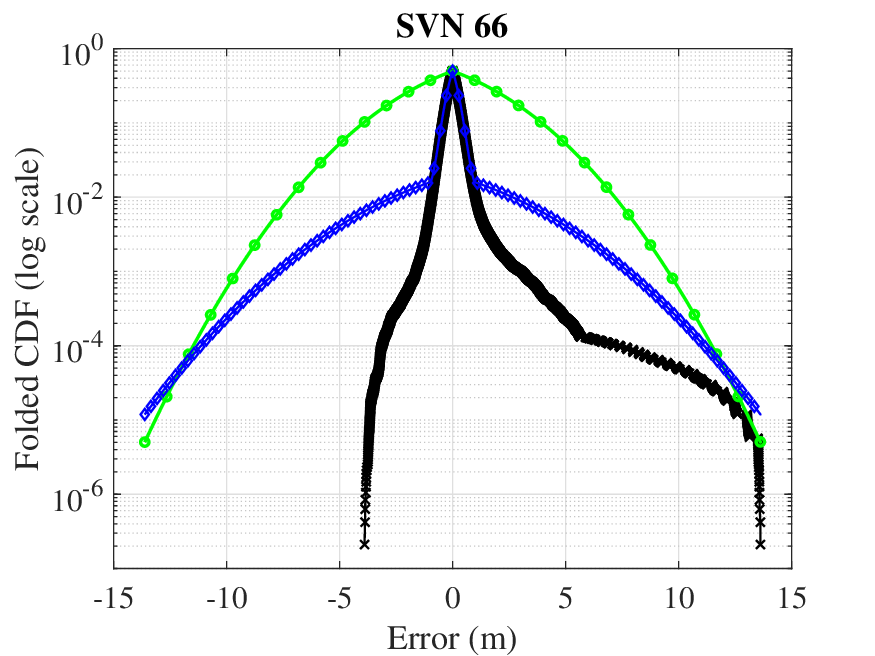}}
    \subfloat[]{%
\includegraphics[width=55mm]{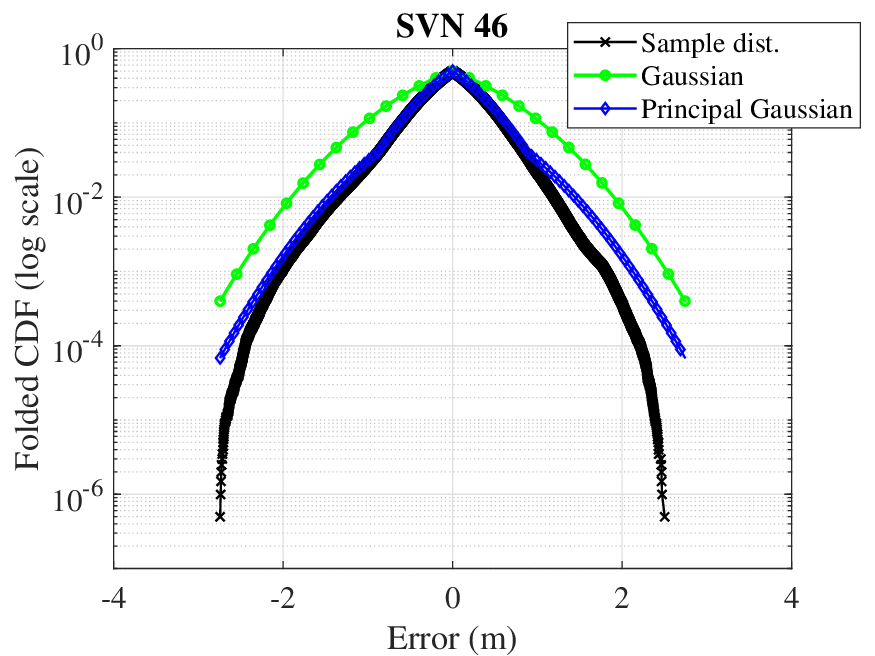}}
    \\
   \subfloat[]{%
\includegraphics[width=55mm]{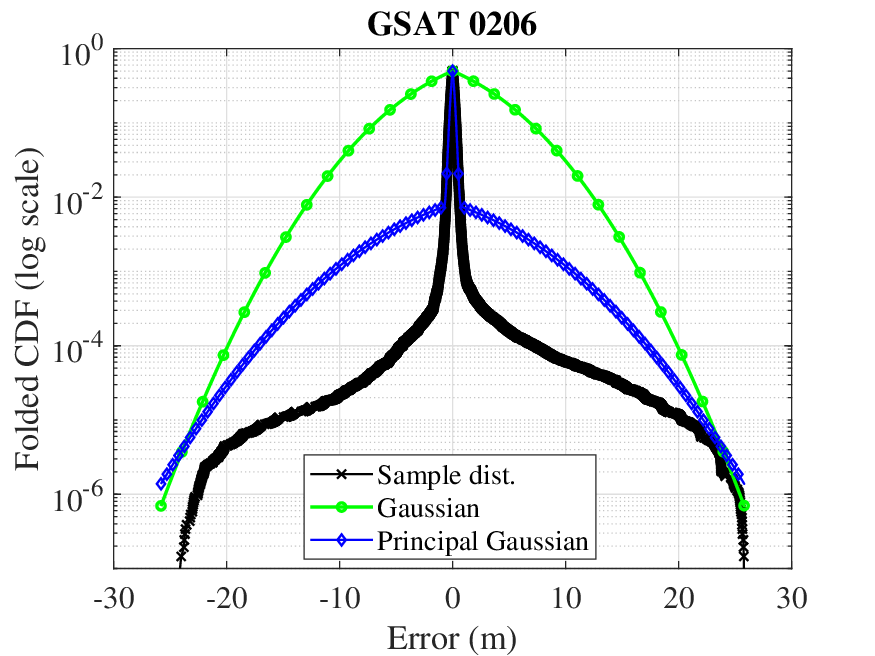}}
   \subfloat[]{%
\includegraphics[width=55mm]{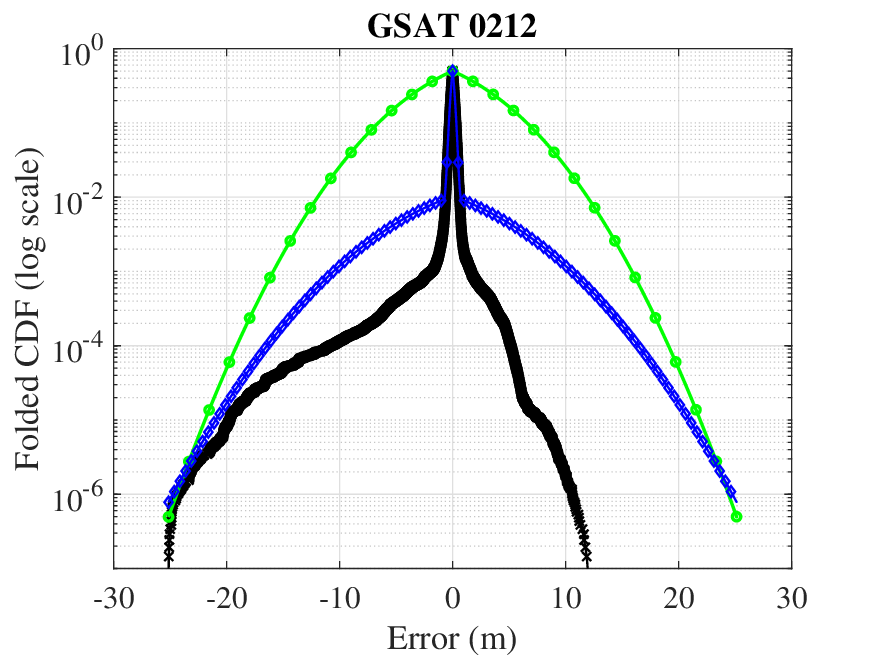}}
       \hfill
\caption{The folded CDF of SISRE and its bounding results for (a) GPS satellite SVN63; (b) GPS satellite SVN66; (c) GPS satellite SVN46;  (d) Galileo satellite GSAT0206; and (e) Galileo satellite GSAT0212.}
  \label{fig:bound_result}
\end{figure}

Three categories of SISRE distributions have been identified in \ref{app:SISRE_performance}.
For each error type, we select one typical satellite from each constellation for detailed analysis.
For two-side heavy-tailed cases, GPS satellite SVN63 (Fig.
\ref{fig:bound_result}a) and Galileo satellite GSAT0206 (Fig.
\ref{fig:bound_result}c) are analyzed.
The SISRE of both satellites demonstrates a significant heavy-tailed phenomenon, with GSAT0206 exhibiting a narrower core (majority of errors within ±2 m) compared to SVN63 (±5 m) but a substantially larger maximum absolute error (26 m vs.
15 m), indicative of heavier tails.
The PGO consistently outperforms the Gaussian overbound, achieving tighter bounds in both the core and tail regions.
For one-side heavy-tailed cases, GPS satellite SVN66 (right-side heavy tail, Fig.
\ref{fig:bound_result}b) and Galileo satellite GSAT0212 (left-side heavy tail, Fig.
\ref{fig:bound_result}d) are examined.
The Gaussian overbound leads to loose bounds on the light-tailed side, whereas the PGO maintains tighter bounds across all error magnitudes.
Finally, Gaussian-like cases are analyzed using solely GPS satellite SVN46 (Fig.
\ref{fig:bound_result}e), as no Galileo satellites exhibit this behavior.
As can be seen, both overbounding methods produce similar results, with negligible differences in bounding performance.
In such cases, the Gaussian overbound is recommended due to its simplicity.
These findings underscore the superiority of PGO for heavy-tailed error distributions while advocating Gaussian overbounding for lighter-tailed scenarios, thereby balancing integrity assurance and computational efficiency.
\par

Tables \ref{tb_GPS_bound} and \ref{tb_Galileo_bound} give the bounding parameters of the Gaussian overbound and the PGO for GPS and Galileo SISRE, respectively.
The 1-sigma Gaussian overbound of GPS SISRE varies significantly, with an average of \SI{1.67}{\meter}.
This is because the SISRE of some GPS satellites exhibits heavy-tailed properties while the others have Gaussian-like behavior, as revealed in \ref{app:SISRE_performance}.
This difference is also reflected in the PGO parameters, where the heavy-tailed SISRE featured with a large gap between $\sigma_1$ and $\sigma_2$, and the Gaussian-liked SISRE has a smaller deviation between $\sigma_1$ and $\sigma_2$.\par

For the Galileo satellites, the 1-sigma Gaussian overbound of SISRE has a smaller variation, with an average of \SI{5.58}{\meter}.
This value aligns closely with the Galileo broadcast User Range Accuracy (URA) parameter, $\sigma_{URA}=\SI{6}{\meter}$, as defined in Galileo Open Service Service Definition Document (OS-SDD) \citep{EU2023Galileo-OS-SDD}.
Since the SISRE of all Galileo satellites exhibits significant heavy-tailed properties, the Galileo broadcast URA parameter is likely to provide an extremely conservative bound for the SISRE.
For the PGO parameters, all Galileo satellites exhibit a high consistency, where $\sigma_2$ is significantly larger than $\sigma_1$, and $p_1$ is larger than 0.98.

\begin{table}[!ht]
\fontsize{8}{7}\selectfont
\caption{\centering Parameters of the Gaussian overbound and the Principal Gaussian overbound of SISRE for each GPS satellite from 1/1/2020 to 12/31/2022. The mean and standard deviation of SISRE are also displayed. }
\label{tb_GPS_bound}
\centering
\begin{threeparttable}
\begin{tblr}{
  width = \linewidth,
  colspec = {Q[200]Q[140]Q[220]Q[200]Q[220]Q[200]Q[200]Q[200]Q[200]},
  cell{1}{3} = {c=2}{c},
  cell{1}{6} = {c=4}{c},
  vline{1,3,5,6,10} = {1-32}{},
  hline{1,3,33} = {-}{},
}
 &  & SISRE &  & Gaussian & PGO &  & &  \\
SVN & Type\tnote{1} & mean (cm) & std (cm) & $\sigma$ (m)    & $\sigma_1$ (m)   & $\sigma_2$ (m)   & $p_1$       & $x_{rp}$ (m)     \\
SVN41 & O & -3.45 & 42.51  & 1.136 & 0.403 & 1.343 & 0.918 & 0.948  \\
SVN43 & T & 1.09  & 52.02  & 1.113 & 0.432 & 1.195 & 0.762 & 0.906  \\
SVN44 & T & -2.3  & 131    & 4.052 & 0.595 & 4.425 & 0.628 & 1.103  \\
SVN45 & O & -4.67 & 42.82  & 1.778 & 0.425 & 2.226 & 0.955 & 1.157  \\
SVN46 & G & -2.44 & 46.16  & 0.818 & 0.413 & 0.78  & 0.787 & 0.884  \\
SVN47 & G & -6.14 & 36.59  & 0.521 & 0.351 & 0.612 & 0.861 & 0.835  \\
SVN48 & G & -2.97 & 55.04  & 0.78  & 0.414 & 0.804 & 0.535 & 0.611  \\
SVN50 & G & -1.99 & 40.48  & 0.574 & 0.411 & 0.691 & 0.977 & 1.493  \\
SVN51 & O & 1.33  & 38.7   & 2.518 & 0.385 & 3.211 & 0.973 & 1.042  \\
SVN52 & G & -4.44 & 49.91  & 0.703 & 0.427 & 0.765 & 0.716 & 0.893  \\
SVN53 & O & -4.49 & 80.98  & 2.245 & 0.54  & 2.426 & 0.624 & 1.074  \\
SVN55 & G & 1.08  & 34.04  & 0.873 & 0.31  & 0.998 & 0.891 & 0.763  \\
SVN56 & G & -4.66 & 37.62  & 0.68  & 0.382 & 0.815 & 0.956 & 1.12   \\
SVN57 & O & -0.96 & 62.42  & 1.08  & 0.471 & 1.333 & 0.806 & 0.79   \\
SVN58 & O & -0.7  & 40.14  & 2.998 & 0.372 & 3.998 & 0.983 & 1.136  \\
SVN59 & G & 2.2   & 35.15  & 0.616 & 0.297 & 0.544 & 0.783 & 0.667  \\
SVN61 & T & -2.55 & 40.05  & 0.753 & 0.321 & 0.837 & 0.788 & 0.684  \\
SVN62 & O & 1.27  & 35.9   & 0.694 & 0.355 & 0.835 & 0.961 & 1.046  \\
SVN63 & T & 3.03  & 46.34  & 3.487 & 0.419 & 4.425 & 0.97  & 1.073  \\
SVN64 & O & 0.38  & 38.94  & 1.495 & 0.39  & 2.05  & 0.985 & 1.155  \\
SVN65 & T & 3.15  & 95.6   & 3.57  & 0.353 & 3.901 & 0.574 & 0.669  \\
SVN66 & O & -1.13 & 39.58  & 3.084 & 0.363 & 3.968 & 0.97  & 0.963  \\
SVN67 & G & -1.51 & 33.14  & 0.54  & 0.292 & 0.6   & 0.84  & 0.649  \\
SVN68 & T & 0.54  & 35.41  & 0.977 & 0.302 & 1.17  & 0.928 & 0.707  \\
SVN69 & T & 4.35  & 65.74  & 3.302 & 0.468 & 3.908 & 0.894 & 1.034  \\
SVN70 & T & 0.86  & 32.3   & 2.303 & 0.308 & 2.959 & 0.965 & 0.821  \\
SVN71 & T & 1.65  & 36.05  & 0.934 & 0.341 & 1.112 & 0.92  & 0.832  \\
SVN72 & G & -4.01 & 121.97 & 1.548 & 1.005 & 1.441 & 0.548 & 0.872  \\
SVN73 & T & -6.99 & 62.01  & 3.68  & 0.521 & 4.11  & 0.842 & 1.154  \\
SVN74 & O & 0.62  & 32.36  & 1.287 & 0.31  & 1.602 & 0.973 & 0.839
\end{tblr}
\begin{tablenotes}
\item[1] ``T": Two-side heavy-tailed; ``O": One-side heavy-tailed; ``G": Gaussian-liked.
\end{tablenotes}
\end{threeparttable}
\end{table}

\newpage
\begin{table}[!ht]
\fontsize{8}{7}\selectfont
\caption{\centering Parameters of the Gaussian overbound and the Principal Gaussian overbound of SISRE for each Galileo satellite from 1/1/2020 to 12/31/2022. The mean and standard deviation of SISRE are also displayed.}
\label{tb_Galileo_bound}
\centering
\begin{threeparttable}
\begin{tblr}{
  width = \linewidth,
  colspec = {Q[250]Q[150]Q[220]Q[200]Q[200]Q[200]Q[200]Q[200]Q[200]},
  cell{1}{3} = {c=2}{c},
  cell{1}{6} = {c=4}{c},
  vline{1,3,5,6,10} = {1-32}{},
  hline{1,3,27} = {-}{},
}
 &  & SISRE &  & Gaussian & PGO &  & &  \\
SVN & Type\tnote{1} & mean (cm) & std (cm) & $\sigma$ (m)    & $\sigma_1$ (m)   & $\sigma_2$ (m)   & $p_1$       & $x_{rp}$ (m)     \\
GSAT0101 & T & -2.11 & 48.13 & 5.967 & 0.292 & 7.717 & 0.985 & 0.79   \\
GSAT0102 & T & -2.66 & 37.72 & 5.758 & 0.311 & 7.662 & 0.984 & 0.909  \\
GSAT0103 & T & -1.5  & 57.17 & 6.098 & 0.289 & 7.43  & 0.98  & 0.752  \\
GSAT0203 & T & -4.69 & 45.92 & 5.89  & 0.338 & 8.278 & 0.986 & 0.967  \\
GSAT0205 & O & -1.34 & 27.12 & 2.333 & 0.229 & 2.867 & 0.984 & 0.68   \\
GSAT0206 & T & -1.38 & 27.76 & 5.346 & 0.236 & 6.859 & 0.986 & 0.717  \\
GSAT0207 & O & -1.48 & 29.75 & 5.724 & 0.256 & 7.188 & 0.983 & 0.758  \\
GSAT0208 & T & -1.12 & 29.3  & 5.687 & 0.246 & 7.144 & 0.985 & 0.74   \\
GSAT0209 & T & -0.81 & 27.3  & 5.423 & 0.232 & 7.245 & 0.986 & 0.682  \\
GSAT0210 & T & -0.36 & 82.48 & 5.714 & 0.23  & 8.783 & 0.98  & 0.57   \\
GSAT0211 & O & -1.32 & 30.73 & 6.197 & 0.234 & 7.809 & 0.984 & 0.715  \\
GSAT0212 & O & -0.96 & 32.05 & 5.136 & 0.25  & 6.351 & 0.983 & 0.725  \\
GSAT0213 & T & -0.3  & 29.95 & 5.97  & 0.251 & 8.416 & 0.984 & 0.691  \\
GSAT0214 & T & -0.55 & 29.92 & 5.561 & 0.238 & 6.926 & 0.983 & 0.693  \\
GSAT0215 & T & -0.31 & 33.8  & 5.619 & 0.238 & 7.483 & 0.985 & 0.694  \\
GSAT0216 & T & -0.96 & 27.89 & 7.383 & 0.229 & 9.264 & 0.983 & 0.698  \\
GSAT0217 & T & -1.23 & 27.23 & 5.518 & 0.228 & 7.16  & 0.986 & 0.673  \\
GSAT0218 & T & -0.87 & 27.81 & 5.598 & 0.229 & 7.031 & 0.983 & 0.676  \\
GSAT0219 & T & -1.3  & 43.8  & 6.155 & 0.28  & 7.761 & 0.986 & 0.795  \\
GSAT0220 & T & 1.87  & 31.99 & 5     & 0.297 & 6.404 & 0.985 & 0.877  \\
GSAT0221 & T & -2.09 & 31.33 & 5.266 & 0.269 & 6.579 & 0.986 & 0.799  \\
GSAT0222 & T & -1.63 & 31.92 & 5.332 & 0.259 & 6.663 & 0.98  & 0.723  \\
GSAT0223 & O & -1.1  & 35.24 & 5.521 & 0.288 & 7.3   & 0.988 & 0.895  \\
GSAT0224 & O & -0.8  & 35.56 & 5.644 & 0.275 & 7.458 & 0.987 & 0.86
\end{tblr}
\begin{tablenotes}
\item[1] ``T": Two-side heavy-tailed; ``O": One-side heavy-tailed.
\end{tablenotes}
\end{threeparttable}
\end{table}

\clearpage
\begin{review}
\section{Additional Results}
In this section, we compare the performance of the \textcolor{black}{SS ARAIM} and jackknife ARAIM both with non-Gaussian nominal error models. 
Fig. \ref{fig:SF_PL_RealSimu_PGO} shows the result in the single-constellation setting. As can be seen, the map of 99.5 percentile of the VPL over the course of a day of both algorithms is exactly the same. The same results are also shown in the dual-constellation setting in Fig. \ref{fig:MF_PL_RealSimu_PGO}. 

\begin{figure}[!htb]
  \centering
\subfloat[~~~~]{%
\includegraphics[width=75mm]{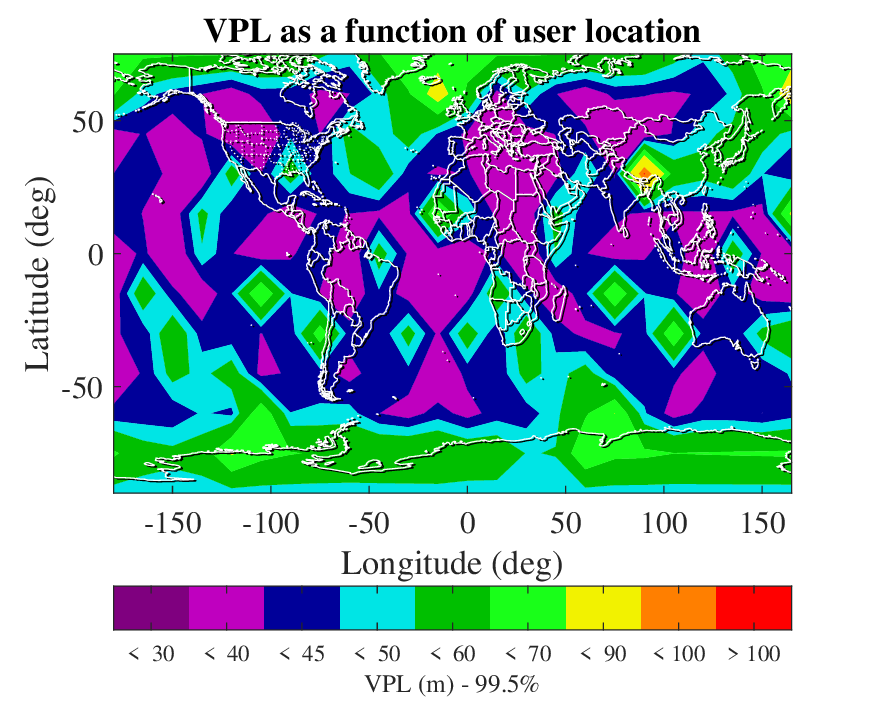}}
  \hfill
\subfloat[~~~~]{%
\includegraphics[width=75mm]{World_RealSimu_SF_PL_PGO_JK.eps}}
\caption{99.5 percentile of the VPL over the course of the day yielded by (a) the \textcolor{black}{SS ARAIM} and (b) the proposed jackknife ARAIM for the single constellation. Both the \textcolor{black}{SS ARAIM} and the proposed Jackknife ARAIM use the PGO for code IF combination nominal errors.}
\label{fig:SF_PL_RealSimu_PGO}
\end{figure}

\begin{figure}[!htb]
  \centering
\subfloat[~~~~]{%
\includegraphics[width=75mm]{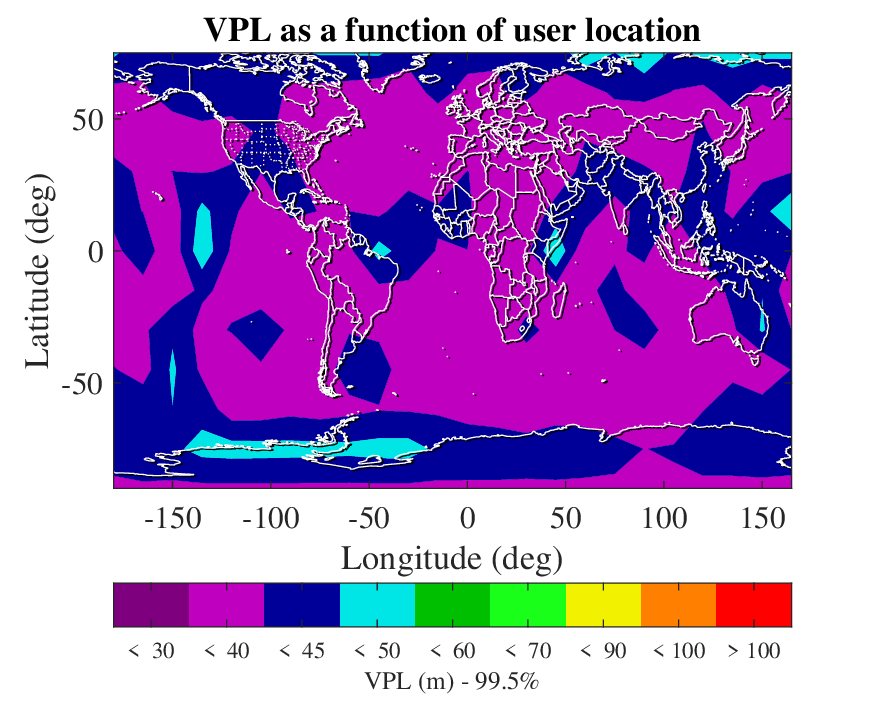}}
  \hfill
\subfloat[~~~~]{%
\includegraphics[width=75mm]{World_RealSimu_MF_PL_PGO_JK.eps}}
\caption{99.5 percentile of the VPL over the course of the day yielded by (a) the \textcolor{black}{SS ARAIM} and (b) the proposed Jackknife ARAIM for the dual constellation. Both the \textcolor{black}{SS ARAIM} and the proposed Jackknife ARAIM use the PGO for code IF combination nominal errors.}
\label{fig:MF_PL_RealSimu_PGO}
\end{figure}
\end{review}
\clearpage

\printbibliography[title=References]

\end{document}